\documentclass[fleqn,usenatbib]{mnras}

\usepackage{amssymb}

\usepackage{newtxmath}
\usepackage{multicol}

\usepackage[T1]{fontenc}
\usepackage[normalem]{ulem}
\DeclareRobustCommand{\VAN}[3]{#2}
\let\VANthebibliography\thebibliography
\def\thebibliography{\DeclareRobustCommand{\VAN}[3]{##3}\VANthebibliography}

\usepackage{graphicx}	
\usepackage{amsmath}	
\usepackage{amssymb}	
\usepackage{bm}

\title[Spectral Separation of SGWBs : Modulated foreground]{Spectral separation of the stochastic gravitational-wave background for \textit{LISA} in the context of a modulated Galactic foreground}

\author[G. Boileau et al.]{
Guillaume Boileau,$^{1}$\thanks{E-mail: guillaume.boileau@oca.eu}
Astrid Lamberts,$^{1,2}$
Nelson Christensen,$^{1}$ 
Neil J. Cornish,$^{3}$
and Renate Meyer$^{4}$
\\
$^{1}$Artemis, Observatoire de la Côte d'Azur, Université Côte d'Azur, CNRS, CS 34229, F-06304 Nice Cedex 4, France\\
$^{2}$Laboratoire Lagrange, Observatoire de la Côte d'Azur, Université Côte d'Azur, CNRS , France\\
$^{3}$eXtreme Gravity Institute, Department of Physics,Montana State University, Bozeman, Montana 59717, USA\\
$^{4}$ Department of Statistics, University of Auckland, Auckland, New Zealand
}

\date{Accepted 07 September 2021. Received 10 May 2021; in original form 20 September 2021}

\pubyear{2021}

\begin{document}
\label{firstpage}
\pagerange{\pageref{firstpage}--\pageref{lastpage}}
\maketitle

\begin{abstract}
Within its observational band the Laser Interferometer Space Antenna, \textit{LISA}, will simultaneously observe orbital modulated waveforms from Galactic white dwarf binaries, a binary black hole produced gravitational-wave background, and potentially a cosmologically created stochastic gravitational-wave background (SGWB). The overwhelming majority of stars end their lives as white dwarfs, making them very numerous in the Milky Way. 
We simulate Galactic white dwarf binary gravitational-wave emission based on distributions from various mock
catalogs and determine a complex waveform from the Galactic foreground with $3.5 \times 10^{7}$ binaries. We describe the effects from the Galactic binary distribution population across mass, position within the Galaxy, core type, and orbital frequency distribution. 
We generate the modulated Galactic white dwarf signal detected by \textit{LISA} due to its orbital motion, and present a data analysis strategy to address it. The Fisher Information and Markov Chain Monte Carlo methods give an estimation of the \textit{LISA} noise and the parameters for the different signal classes. We estimate the detectable limits for the future \textit{LISA} observation of the SGWB in the spectral domain with the three \textit{LISA} channels $A$, $E$, and $T$. 
We simultaneously estimate the Galactic foreground, the astrophysical and cosmological backgrounds. Assuming the expected astrophysical background and a Galactic foreground, a cosmological background energy density of around $\Omega_{GW,cosmo} \approx 8 \times 10^{-13}$ could be detected by \textit{LISA}. \textit{LISA} will either detect a cosmologically produced SGWB, or set a limit that will have important consequences. 

\end{abstract}




\section{Introduction }

The latest \textit{Gaia} data release, the Early Data Release 3 (EDR3), was recently presented \citep{2020arXiv201202061G}. \textit{Gaia} is an astrometry mission, it measures with great precision the position, parallax, and movement of hundreds of millions of stars in our Galaxy. Moreover, with its spectrometer, it is possible to know the type of most of the stars observed.  This is the most accurate stellar map to date giving the position, the luminosity, and the spectrum of more than $ 1.8 \ \times 10^9 $ stars. Among these, $10^5$ white dwarfs (WDs) have been observed and well separated from other stars in the Hertzsprung–Russell diagram \citep{2019MNRAS.482.4570G}.  Stars with an initial mass between 0.9 and 8 solar mass ($M_{\odot}$) will become a WD within a Hubble time. 
 This implies that 97\% of stars in the Galaxy will finish as a WD  \citep{2009JPhCS.172a2004N,2001PASP..113..409F}, resulting in 10 to 50 billion WDs in the Milky Way. For 50 billion WDs, we use the density of $5 \times 10^{-3} \text{ pc}^{-3}$. In \cite{2001JRASC..95...32L}, the distribution  of the star class differences in our galaxy is given. The calculation is based on the number stars of each type in a volume of $10^4 \text{ pc}^3$ around the Sun.

WDs are  stellar core remnants, have typical radii around 10,000 km and masses between $ 0.25 M_{\odot} $ for the He WDs and up to the Chandrasekhar mass $ 1.4 M_{\odot} $, which makes them compact objects \citep{10.1093/mnras/91.5.456}. There are a significant number of WDs that form double WDs (DWD) \citep{2001A&A...365..491N}. Ultra-compact WD binaries with a short orbital period, from a few hours down to a few minutes, can have a significant electromagnetic (EM) signal, making them observable. Among these are cataclysmic variable (CV) systems ~\citep{2020ApJ...891...45K} made of a  WD and a companion star, which transfers part of its mass after having filled its Roche lobe. When matter falls towards the WD, there is a strong periodic emission of UV and sometimes X-rays \citep{warner_1995}. Such interacting binaries are possible progenitors of Type Ia supernova \citep{1984ApJ...277..355W,doi:10.1146/annurev.astro.38.1.191}.

DWDs in our Galaxy are sources of gravitational waves (GW) that will be detectable with \textit{LISA} \citep{2017arXiv170200786A}, the future space mission of the European Space Agency (ESA) whose objective is to detect low frequency GW from space. Its observational band is a great source for understanding the astrophysical properties of our Galaxy and the DWD population.

 \textit{Gaia} DR2 provided astrometry for some WDs. We cannot yet distinguish individual DWDs binaries, but for some known binaries the estimate of their GW emission could be refined. 
\textit{Gaia} DR4 will identify many sources that will be detectable by \textit{LISA} with a signal-to-noise ratio (SNR) larger than five~\citep{2018MNRAS.480..302K}; note also that many systems which can be found by \textit{Gaia} will not have a larger SNR for \textit{LISA}~\citep{2017MNRAS.470.1894K,2019MNRAS.482.4570G}.
From the light curves, it will be possible to extract an estimate of the DWD population that is detectable by \textit{LISA}~\citep{2018MNRAS.480.3942H}. The GW emission will be at frequencies less than one $10^{-3}$ Hz. The study and measurement of these systems are some of the key science goals  of the \textit{LISA} mission. In addition, there are binary systems known as `verification binaries' \citep{2018MNRAS.480..302K}. 
For example, recently the Zwicky Transient Facility (ZTF) has measured a DWD with an  orbital period measured at 7 minutes \citep{2019Natur.571..528B}, which corresponds to a GW emission of $\simeq 3 \times 10^{-2}$ Hz. Well studied systems like this can be used to verify the \textit{LISA} performance, acting as a way to confirm the sensitivity of \textit{LISA}.

 The band $10^{-5}$ to $10^{-4}$ Hz, if usable with \textit{LISA} data, would be important for the detection and separation of the stochastic gravitational wave backgrounds (SGWBs). The high frequency (up to $0.1$ Hz) is dominated by the LISA noise, so it is difficult to separate a SGWB from this noise. However, the high frequency data does provide some important information about the \textit{LISA} noise.
The goal of the \textit{LISA} mission~\citep{2017arXiv170200786A}  is to detect GWs in the $[10^{-4}, 0.1]$ Hz frequency band, possibly extendable to $[10^{-5}, 1]$ Hz. This corresponds to orbital periods between 12 seconds and 15 days. 

The study of the population of DWDs is an important goal for the \textit{LISA} mission. \textit{LISA} is being led by the ESA, with participation from NASA. The launch is currently planned for 2034, with at least 4 years of observations, possibly extended to 10 years. The \textit{LISA} constellation will consist of three spacecrafts separated from one another by $L = 2.5 \times 10^9$ m.
There are many GW signals expected to be detected in the \textit{LISA}  band. 
Galactic sources will be significant for \textit{LISA}, for example from DWD systems~\citep{Nelemans_2001, PhysRevD.76.083006, Ruiter_2010, 2014PhRvD..89b2001A, 10.1093/mnras/stz2834, 2017PASA...34...58E, Hernandez_2020}.
The stochastic GW signal from the DWDs, or the Galactic foreground, is anisotropic and the representation of its energy density is not a simple power law \citep{PhysRevD.64.121501}.
Many studies have addressed populations of DWDs in our Galaxy and their detectability in the \textit{LISA} band. 
\citet{2012ApJ...758..131N} computes the stochastic signal of Galactic origin according to different DWD models. \citet{2020ApJ...901....4B} presents a method of calculating the Galactic foreground and discusses the power distribution and resolvability by \textit{LISA} as a function of distance to the source.  \citeauthor{2014PhRvD..89b2001A} introduce the calculation of the orbitally induced modulation of the Galactic foreground in the context of detecting a stochastic GW background (SGWB) of cosmological origin \citep{2002PhRvD..65b2004C,2014PhRvD..89b2001A,2010PhRvD..82b2002A}. \citet{2020A&A...638A.153K} and \citet{2020ApJ...894L..15R} explore the possibility of observing DWDs in satellite Galaxies. 

The SGWB is the superposition of the large number of independent GW sources~\citep{Romano2017,Christensen_2018}. 
We can distinguish three types of SGWB in the \textit{LISA} band, depending on their origin. 
The most important  by its amplitude is the \textbf{Galactic foreground} produced by the DWDs in our Galaxy. We note that it mainly consists of DWDs, but there are other types of galactic sources which contribute to the galactic foreground, CVs, Stripped stars or WD+M-dwarfs for example. We simulate this foreground with mock catalogs of DWDs.  The second source is the background from extragalactic binary black holes (BBH) and binary neutron stars (BNS) throughout the universe which we call the \textbf{astrophysical background}. This background is also present in the LIGO and Virgo band (roughly between 20 and 1000 Hz) and by considering this background as a power law, it is possible to extrapolate this background in the \textit{LISA} band \citep{2019ApJ...871...97C,PhysRevLett.116.131102}. 

It is also possible to use binary population synthesis models to construct an astrophysical population of BBH and BNS and to predict the associated SGWB \citep{2020arXiv200804890P}. Finally, the cosmological background~\citep{Caprini:2018mtu} denotes the stochastic background coming from the primordial processes such as inflation, phase transitions or cosmic strings \citep{Sakellariadou:2009ev,CHANG2020100604}. The cosmological SGWB originates in the early universe \citep{1995PhRvD..52.2083M,GarciaBellido:2007dg}, and its measurement may allow for the estimation of parameters related to the physical processes at this initial period \citep{2020arXiv200704241C}. 

The cosmic string study of \citet{Auclair_2020} describes the possibility for LISA of detecting a minimum string tension around $G\mu \leq 10^{-17}$, which corresponds to a plateau around $\Omega_{GW}^{plateau} \simeq 5 \times 10^{-12}$. The review of \citet{Caprini_2018} states that it will be possible with LISA to measure a SGWB from a phase transition with $ \Omega_{GW} \simeq 10^{-13} $; there is much uncertainty as to the existence of a phase transition SGWB source in the LISA observation band, let alone its signal strength. The review by \citet{Christensen_2018} describes a limit of detectability with LISA of $\Omega_{GW}(f \simeq 10^{-3} \text{ Hz}) \simeq 5 \times 10^{-13}$ for a standard inflation produced SGWB. The SGWB level from inflation is probably $\Omega_{GW}(f \simeq 10^{-3} \text{ Hz}) \approx 10^{-15}$, or lower. See these reviews~\citep{Caprini_2018,Christensen_2018} for descriptions of other possible cosmogically produced SGWBs.
These studies cited above correspond to an ideal case of a cosmological source and LISA noise, and have not included the effects of a galactic foreground nor an astrophysical background (as we do in this paper). 

Many recent studies explore avenues to detect a cosmological SGWB in the presence of an astrophysical SGWB, and a brief review is presented in our previous study~\citep{2020arXiv201105055B}. The goal of this present paper is to address the possibility for \textit{LISA} to observe a SGWB of cosmological origin  in the presence of other stochastic signals. Our previous study \citep{2020arXiv201105055B} addressed the detectability by \textit{LISA} of a cosmologically produced SGWB in the presence of different levels of a BBH produced astrophysical background.  This study did not consider the Galactic foreground, but did demonstrate the utility of using Bayesian parameter estimation methods for spectral separability. Adding the Galactic foreground  is the goal of the study given in this paper. We present an algorithm to calculate the parameters associated with the Galactic foreground seen by \textit{LISA}. The calculation is aided by the fact that the Galactic foreground experiences a modulation over a year as the \textit{LISA} constellation orbits the sun and changes its orientation with respect to the Galactic center. We use a mock Galactic DWD catalog as the input for calculating the GW foreground. We highlight the quantities which introduce the most variation in the energy spectrum of the Galactic foreground and use that knowledge to predict its form. We also present a strategy to separate the three stochastic signals (Galactic, astrophysical, and cosmological), as well as the inherent \textit{LISA} detector noise, using a Bayesian strategy \citep{PhysRevD.58.082001,PhysRevD.76.083006} based on an Adaptive Markov chain Monte-Carlo (A-MCMC) algorithm. 

The remainder of this paper is organized as follows.
In  Section~\ref{sc:catalog}, we present the DWD catalog and the production of GWs from binary systems. In Section~\ref{sec:waveform} we calculate the waveform for each binary, and how \textit{LISA} responds to this Galactic foreground. The spectrum of the Galactic DWD foreground is calculated and presented in Section~\ref{sc:SpecSepa}, as well as a brief summary of the A-MCMC methods. 
Section~\ref{sc:cov} gives a description of the \textit{LISA} data channels and methods used to describe the \textit{LISA} detector noise. 
Section~\ref{sc:modulation} presents the strategy to identify the Galactic foreground using the information from the orbital modulation of the \textit{LISA} signal. The limits for \textit{LISA} to observe a cosmological SGWB are presented in Section~\ref{sc:cosmolimitation}. The conclusions for our study are given in Section~\ref{sec:conclusion}.

\section{Description of the catalogs of Double White Dwarfs (DWDs)} \label{sc:catalog}
\subsection{Simulation of the DWDs}\label{sc:simuDWD}
\citet{10.1093/mnras/stz2834} provide a catalog of short-period WD binaries producing GWs in \textit{LISA}'s observational frequency band. We refer the reader to this publication for a detailed description of the catalog. 
This simulation of the large population of binaries ($\simeq 3.5 \times 10^{7}$) is based on the "Latte" \citep{10.1093/mnras/stu1738, Wetzel_2016} model of a Milky-Way-like Galaxy from a cosmological simulation in the FIRE project \citep{Hopkins18_FIRE}. The simulation provides a realistic model for the star formation history, metallicity evolution and morphology of the Milky Way, including statistical properties of its satellite population.

It is of course possible to calculate the galactic foreground from other catalogs and compare the effects of the different populations. Here, we also use the MLDC catalog\footnote{\url{https://lisa-ldc.lal.in2p3.fr/}}.  
The Galaxy model is combined with a distribution of DWDs based on a binary population synthesis model \citep{Hurley02_BSE} which naturally produces DWDs with different core compositions depending on initial conditions. Each core composition has a different mass distribution; the He cores are less massive than CO and NeO cores. The formation of CO-CO DWDs typically occurs on timescales shorter than 2 Gyrs while He-He DWDs form on timescales of at least 3 Gyrs. These different delay times result in a distinct distribution of He-He DWDs dominating in the older regions of the Galaxy (thick disk, bulge and halo) and the CO-CO DWDs dominating in regions of more recent star formation (thin disk). 
The simulations converge to parameters similar to those of the Milky Way \citep{2020ApJS..246....6S}.   
The simulation calculates the stellar formation with the position of object in the Galaxy (X,Y, Z), and the metallicity $\mathcal{Z}$ over time; it also 
uses a modified version of the publicly available Binary Star Evolution (BSE) \citep{Hurley02_BSE} to replicate the population of DWDs.

The \textit{LISA} \textit{LDC 1-4} catalog is a Galactic white dwarf binaries population comprising about 30 million systems \citep{Babak_2008}. The catalog contains for each binary the ecliptic latitude and longitude, the amplitude, the frequency, the frequency derivative, the
inclination, and the initial Polarisation. All these parameters respect the distribution given by
\citet{Nelemans_2001}

 The results of the simulation and the mock \textit{LISA} catalog are compatible. The study compares the simulations with what has been observed in our Galaxy. 

\subsection{Comparison of catalogs}\label{sc:catacomp}
The catalog of \citet{10.1093/mnras/stz2834} contains for each binary: the mass of the two stars, $M_1$ for the biggest object and $M_2$ for the smaller; the nature of the core of the star, helium core $He$, carbon-oxygen core $CO$, or neon-oxygen core $NeO$; the orbital frequency of the binary $f_{orb}$; and the Cartesian position in the Galaxy $X,Y,Z$. 

It is straightforward to derive the quantities necessary to describe GW emission from these parameters. The chirp mass is given by:
\begin{equation}
    \mathcal{M}_c = \frac{\left(M_1 M_2\right)^{3/5}}{\left(M_1+M_2\right)^{1/5}}.
\end{equation}
The frequency of the GW emitted by each binary is 
\begin{equation}
\label{eq:f2}
    f_{GW} = 2 f_{Orb} ~ ,
\end{equation}
with $f_{Orb}$ the orbital frequency; we assume that the orbits are circular.
 
The GW frequency derivative is given by 
\begin{equation}
    \dot{f}_{GW} = \left(\frac{G\mathcal{M}_c}{c^3}\right)^{5/3}\frac{96}{5}\pi^{8/3}f_{GW}^{11/3}.
\end{equation}
The distance between the binary and \textit{LISA} (approximating the \textit{LISA} constellation position at the Sun): 
\begin{equation}
    R = \sqrt{(X-X_{\odot})+(Y-Y_{\odot})+(Z-Z_{\odot})}
\end{equation}
with $(X_{\odot},Y_{\odot},Z_{\odot}) = (8.178, 0, 0.659) \text{ kpc}$ the position of the Sun in the Galactic Cartesian coordinates. 

\begin{figure}
    \centering
    \includegraphics[height= 4.5cm]{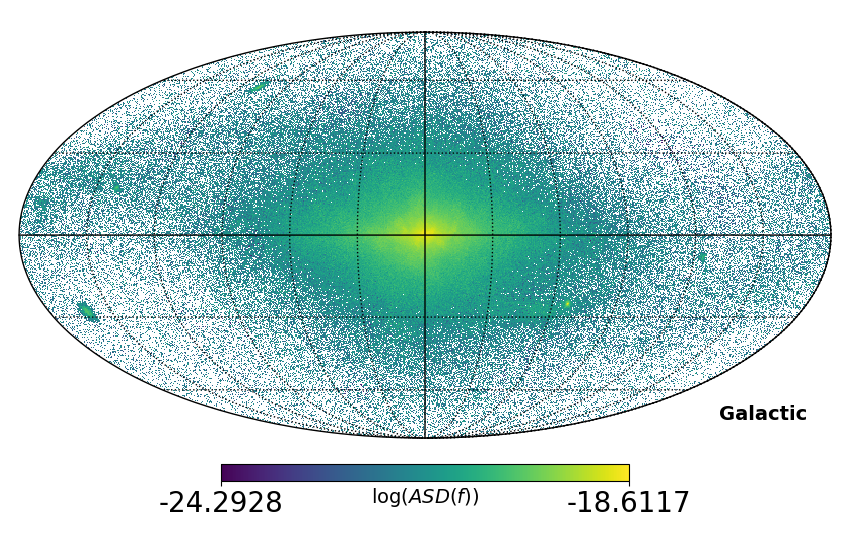}
    \caption{Map of the distribution of the log-amplitude of GW from the Galactic WD binaries for $f_{GW} \ge 1 \times 10^{-5} \ \text{Hz}$. The DWD distribution in position ($X,Y,Z$) is from the simulation of \citet{10.1093/mnras/stz2834}. This map is made with the Galactic coordinates $GLON,GLAT$ with $N_{side}=256$.}
    \label{fig:healpix}
\end{figure}

\begin{figure}
    \centering
    \includegraphics[height= 4.5cm]{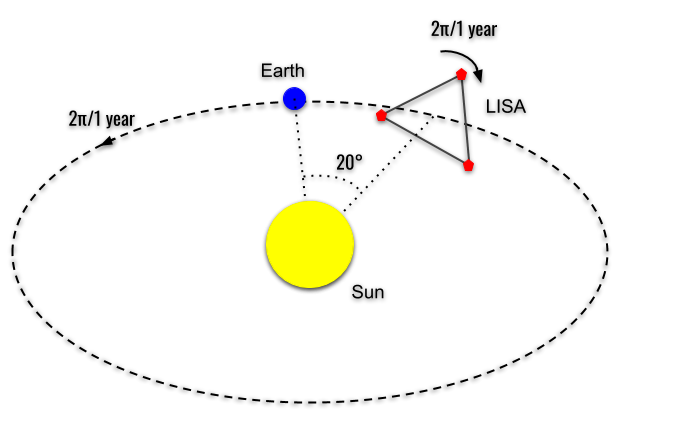}
    \caption{\textit{LISA} constellation of three spacecraft. They will be placed in a heliocentric orbit and form an equilateral triangle of 2.5 million kilometer arm length. Each satellite's distance to another will be measured by laser beams. The constellation's orbit forms an angle of delay of $20^{\circ}$ with respect to that of the Earth's. Just as \textit{LISA} makes an orbit of period 1 year, during this time, it also performs a revolution on itself.}
    \label{fig:LISA_constellation}
\end{figure}

\begin{figure}
    \centering
    \includegraphics[height= 4.5cm]{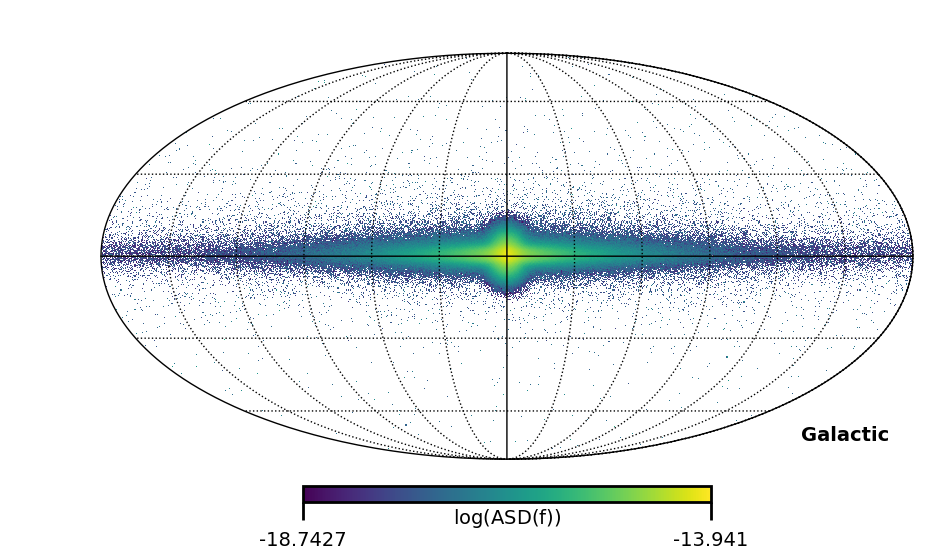}
    \caption{Map of the distribution of the log-amplitude of the GW from the Galactic WD binaries for $f_{GW} \ge 1 \times 10^{-4}$ Hz. The DWD distribution is from the population presented in \citet{Nelemans_2001}.}
    \label{fig:healpix_nelemans}
\end{figure}

For a DWD, according to \citet{PhysRevD.76.083006} we can compute the GW amplitude for an optimally polarized and aligned binary at a distance $R$ as,
\begin{equation} \label{eq:h}
    h_{opt} = 4(\pi)^{2/3} f_{GW}^{2/3}\frac{G^{5/3}}{c^4}\frac{M_c^{5/3}}{R} ~ .
\end{equation}
This is sufficient for mapping the amplitude. We use the healpix \citep{2005ApJ...622..759G} view in Fig.~\ref{fig:healpix}, with $N_{side}=256$ \citep[~see definition $N_{side}$ in][]{2005ApJ...622..759G}. For each pixel on the map we stack the amplitudes of binaries present; Fig.~\ref{fig:healpix} and Fig.~\ref{fig:healpix_nelemans} are  the logarithm of the sky GW amplitude background  of \citet{10.1093/mnras/stz2834} and \citet{Nelemans_2001}, respectively. The \textit{LISA} constellation position is displayed in Fig.~\ref{fig:LISA_constellation}. Fig.~\ref{fig:healpix} is constructed with the positions of the binaries of \citet{10.1093/mnras/stz2834}. As introduced in Sec.~\ref{sc:simuDWD}, this is the result of a simulation having as input the astrophysical parameters, while the figure is constructed with positions independent of the frequencies; the positions follow an exponential distribution simulating the shape of the galaxy \textbf{\citep{doi:10.1111/j.1365-2966.2011.18564.x}}.
For our study we generate GW signals starting at $10^{-5}$ Hz. 

We introduce as well the DWD population from the \textit{LISA} DATA Challenge~\citep{LDC}. \textit{LDC 1-4} uses the Galactic position distribution from \citet{Nelemans_2001}. The Galactic population is symmetrically chosen in a random fashion for the disk and the bulge. The other parameters are also randomly chosen with distributions that address the Galactic birth rate and evolution scenario. 

This model gives a binary rich Galactic center and the arms. 
However, there are few binaries in the rest of the sky. In comparison, the population of \citet{10.1093/mnras/stz2834} (see Fig.~\ref{fig:healpix}) has a distribution closer to our Galaxy. Indeed the simulated Galaxy contains a disk, a bulge, a halo and satellite Galaxies. 
In addition, there is the presence of DWDs all over the sky but with an anisotropy.

\subsection{Amplitude calculation}\label{sc:amplcal}
The GW amplitudes for the two polarisations from a binary are given by:
\begin{equation}
    A_+(M_1,M_2,R,f,\iota) = \frac{2 G^2 M_1 M_2}{c^4 R}\left ( \frac{(\pi f)^2}{G(M_1+M_2)}\right) ^{1/3}(1+\cos^2(\iota)),
\end{equation}
\begin{equation}
    A_{\times}(M_1,M_2,R,f,\iota) = -\frac{4 G^2 M_1 M_2}{c^4 R}\left ( \frac{(\pi f)^2}{G(M_1+M_2)}\right) ^{1/3}{\cos}(\iota).
\end{equation}
In the calculation of \citet{10.1093/mnras/stz2834} there is no inclination for the orbital plane of the binary, $\iota$. We assume the distribution of the inclination will be uniform for $\cos(\iota)$. We integrate the two amplitudes over $\cos(\iota)$:
\begin{equation}
\begin{split}
       \underline{A} &=  \sqrt{\int_{-1}^{1} (A_+(\iota)^2 + A_{\times}(\iota)^2)  \mathrm{d}(\cos(\iota))} \\
  &= \frac{4 G^2 M_1 M_2}{c^4 R}\left ( \frac{(\pi f)^2}{G(M_1+M_2)}\right) ^{1/3} \underline{A}_{\iota}
    \end{split}
\end{equation}
with $\underline{A}_{\iota} = \sqrt{\int_{-1}^{1} ((1+y^2)^2/4 + y^2) \mathrm{d}y} = \sqrt{\frac{8}{5}}$, which gives
\begin{equation}
\begin{split}
\underline{A} &= 4(\pi)^{2/3} f_{GW}^{2/3}\frac{G^{5/3}}{c^4}\frac{M_c^{5/3}}{R} \underline{A}_{\iota}.
\end{split}
\end{equation}
Below, Sec.~\ref{sec:waveform}, we give the response of LISA to both gravitational-wave polarisations, and in deriving the final results for our study we average over $cos(\iota)$.

For the DWD population we can compute the polarisation-averaged $h$. We use Eq. \ref{eq:h} for a binary located at 1 kpc from the \textit{LISA} constellation with an orbital period of one hour and with a chirp mass of 1 solar mass:
\begin{equation}
h = 1.08 \times 10^{-21} \left(\frac{\mathcal{M}_c}{ 1 \text{ M}_{\odot}}\right)^{5/3} \left(\frac{\mathcal{P}_{orb}}{ 1 \text{ hr}}\right)^{-2/3} \left(\frac{R}{ 1 \text{ kpc}}\right)^{-1}
\end{equation}
where $R$ is the distance between \textit{LISA} and the binary in kpc and the orbital period $\mathcal{P}_{orb} = \frac{1}{f_{orb}}$. An orbital period of 1 hour corresponds to an orbital frequency $f_{orb} = 2.8 \times 10^{-4} \text{ Hz}$.   
We define the amplitude spectral density ($ASD$) as:
\begin{equation}\label{eq:ASD}
\sqrt{S_h(f)} = ASD(f) = \frac{h}{ \sqrt{2T_{Obs}}}
\end{equation}
with $T_{Obs} = 4$ years and $S_h(f)$ is the power spectral density of the binary signal \citep[see][Eq. 19]{Robson_2019}.
We can predict the amplitude spectral density for each binary, and compare the population with the \textit{LISA} sensitivity $S_n(f)$ \citep{Robson_2019}:
\begin{equation}
    \begin{split}
        S_n(f) = \Bigg[&1.2 \times 10^{-40} \ \text{Hz}^{-1} \Bigg(1 + \left(\frac{2 \times 10^{-3} \ \text{Hz}}{f}\right)^4 \Bigg) \\
        &+ 9.6 \times 10^{-48}\ \ \text{s}^{-4}\text{Hz}^{-1}\Big(1 + \cos^2\frac{f}{f_{ref}}\Big) \frac{\left(1 + \frac{4 \times 10^{-4} \ \text{Hz}}{f}\right)^2}{(2\pi f)^4} \\
        &\times \left(1 + \frac{f}{\left(8 \times 10^{-4} \ \text{Hz}\right)^4}\right) \left(1 + 0.6\left(\frac{f}{f_{ref}}\right)\right)^2  \Bigg]^{1/2}.
    \end{split}
    \end{equation}

\begin{figure*}
    \centering
    \includegraphics[height= 8cm]{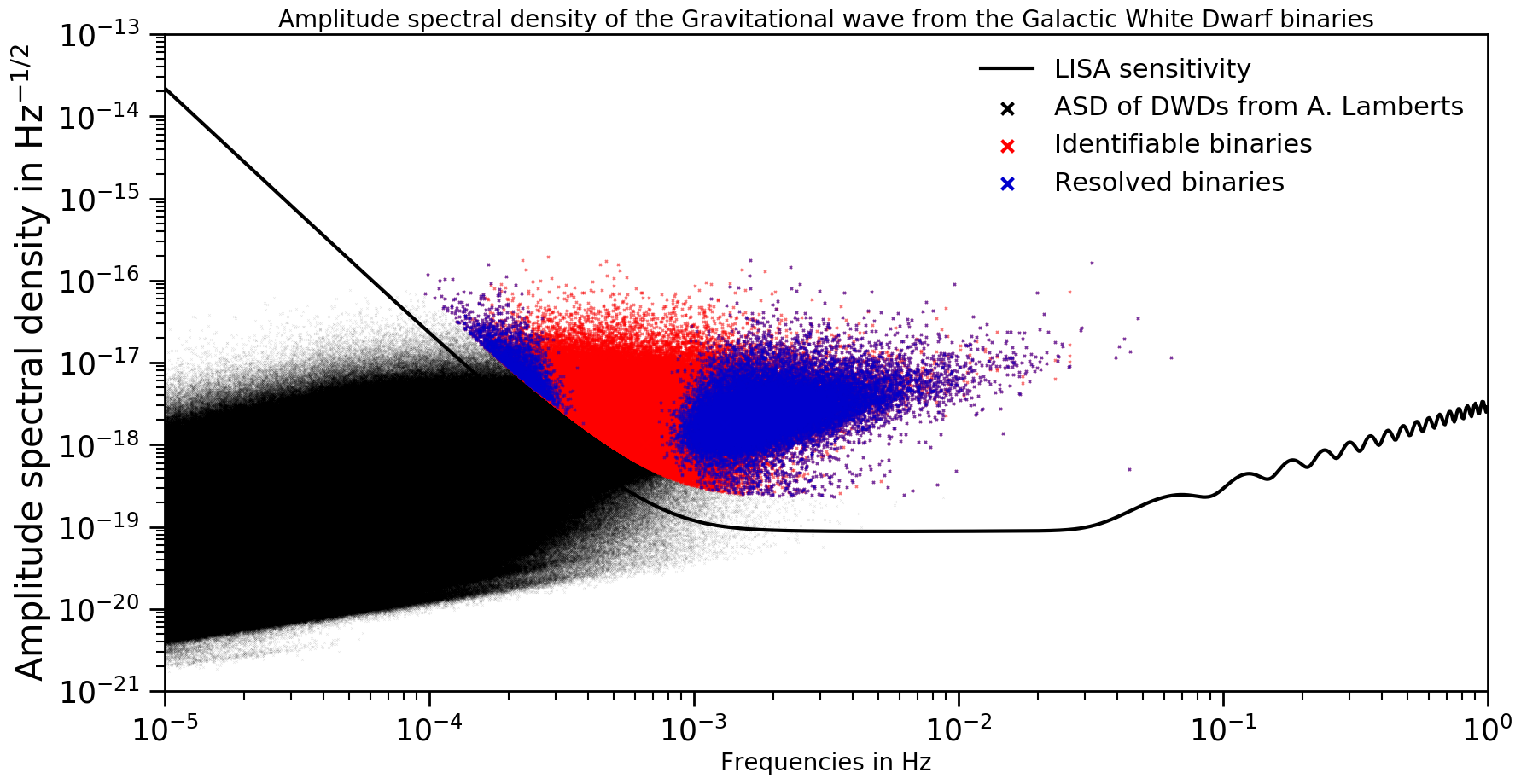}
    \caption{Amplitude spectral density for the GW from the Galactic WD binaries for four years of duration of science time  with the \textit{LISA} strain sensitivity $S_n(f)$ (black line). The black scatterplot displays the totality of binaries from the \citet{10.1093/mnras/stz2834} catalog of DWD. We calculate for each DWD the amplitude spectral density (ASD); see Eq.~\ref{eq:ASD}. In red, we calculate the "identifiable" binaries; these are binaries of $SNR > 7$~\citep{PhysRevD.73.122001}. The blue population corresponds to "resolved" binaries, namely no more than one per \textit{LISA} bin; we calculate for each binary a spectral separation from all the other binaries in comparison to the \textit{LISA} frequency bin size of $\frac{1}{T_{Obs}}$}
    \label{fig:ASD}
\end{figure*}    
In Fig.~\ref{fig:ASD}, the black scatter plot shows the amplitude spectral density $(ASD)$ of all the binaries from \citet{10.1093/mnras/stz2834}; the red dots are the \textit{identifiable} binaries, and the blue dots are {\textit{resolved} binaries}.
In the \textit{LISA} band $f_{GW} \ge 10^{-5} \text{Hz}$, we expect signals from $\sim$ 35 million binaries. We restrict our study to binaries with a GW frequency greater than $10^{-5} \text{Hz}$. Approximately one in a thousand binaries will be resolvable, leaving the large majority of the Galactic binaries unresolved in a stochastic signal. A DWD is \textit{resolved} if is uniquely \textit{identified} in its frequency bin and has SNR  $> 7$~\citep{PhysRevD.73.122001}. The presence of a signal from a DWD can be identified for SNR  $> 7$, but if there is more than one signal per bin it will not be possible to resolve it individually, we refer such binaries as \textit{identifiable}. A resolved source has a frequency difference with any other binary larger than the \textit{LISA} bin $\frac{1}{T_{Obs}}$. The frequency derivative of the gravitational wave is $\dot{f}_{GW} \propto \mathcal{M}_c^{5/3} f_{GW}^{11/3}$. Considering the example at low frequency of $f_{GW} = 0.06$ Hz and $\mathcal{M}_c = 1 M_{\odot}$; this implies $\dot{f}_{GW} = 2.48 \ \times 10^{-11} \ \text{Hz} \ \text{s}^{-1}$.

For a 4 year duration, we have a maximum frequency shift $\dot{f}_{GW}T_{Obs} $ of $0.0003$ Hz; the maximum relative frequency shift $\frac{\dot{f}_{GW}T_{Obs}}{f_{GW}}$ in the catalog is 0.5\%. Hence the orbital GW emission can be considered as monochromatic. We can also calculate the coalescence time 
\begin{equation}
\label{eq:tauc}
\tau_c = \frac{5}{256}\frac{c^5a^4}{G^3M_1M_2(M_1+M_2)} ~ ,
\end{equation}
with $a$ the initial separation between the two WDs, given by Kepler's third law, ${f_{orb}}^2a^3 = G (M_1 + M_2)$. The smallest coalescence time of the population is 23 500 years and the biggest is 26 600 times the age of the Universe.  Resolved binaries are separated in two populations (see Fig.~\ref{fig:ASD}). The blue (identifiable binaries for a \textit{LISA} bin = $\Delta f = \frac{1}{T_{Obs}}$) left population $f_{GW}\in [1\times 10^{-4},2 \times 10^{-4}]$ Hz consists of small binaries (less mass); there a large number of sources at low frequencies. The \textit{LISA} noise is relatively high, so a large number of these binaries are not identifiable. However, there are some resolved binaries because they are located close to \textit{LISA}. In Fig.~\ref{fig:ASD}, the blue right population $f_{GW}\in [7\times 10^{-4},5 \times 10^{-2}]$ Hz is produced by the largest objects in terms of mass, but with a small number of them and a dispersion of amplitudes. The middle part $f_{GW}\in [2\times 10^{-4},7 \times 10^{-4}]$ Hz is where there are many observable binaries, a region where the \textit{LISA} noise is low. Because of a large number of sources, separation in frequency is smaller than the frequency bin size $\frac{1}{T_{Obs}}$.
When \textit{LISA} will be observing galactic binaries there will be only four pieces of information per frequency bin, namely the real and imaginary parts in the $A$ and $E$ channels. As such, when there are more than one binary per two frequency bins, there will be more parameters than data points, and resolution of an individual binary will be challenging. However, recent studies have made progress in characterizing the galactic foreground coming from an astrophysical population of binaries, as in ~\citet{2021arXiv210314598K}.

\begin{figure*}
    \centering
    \includegraphics[height= 10cm]{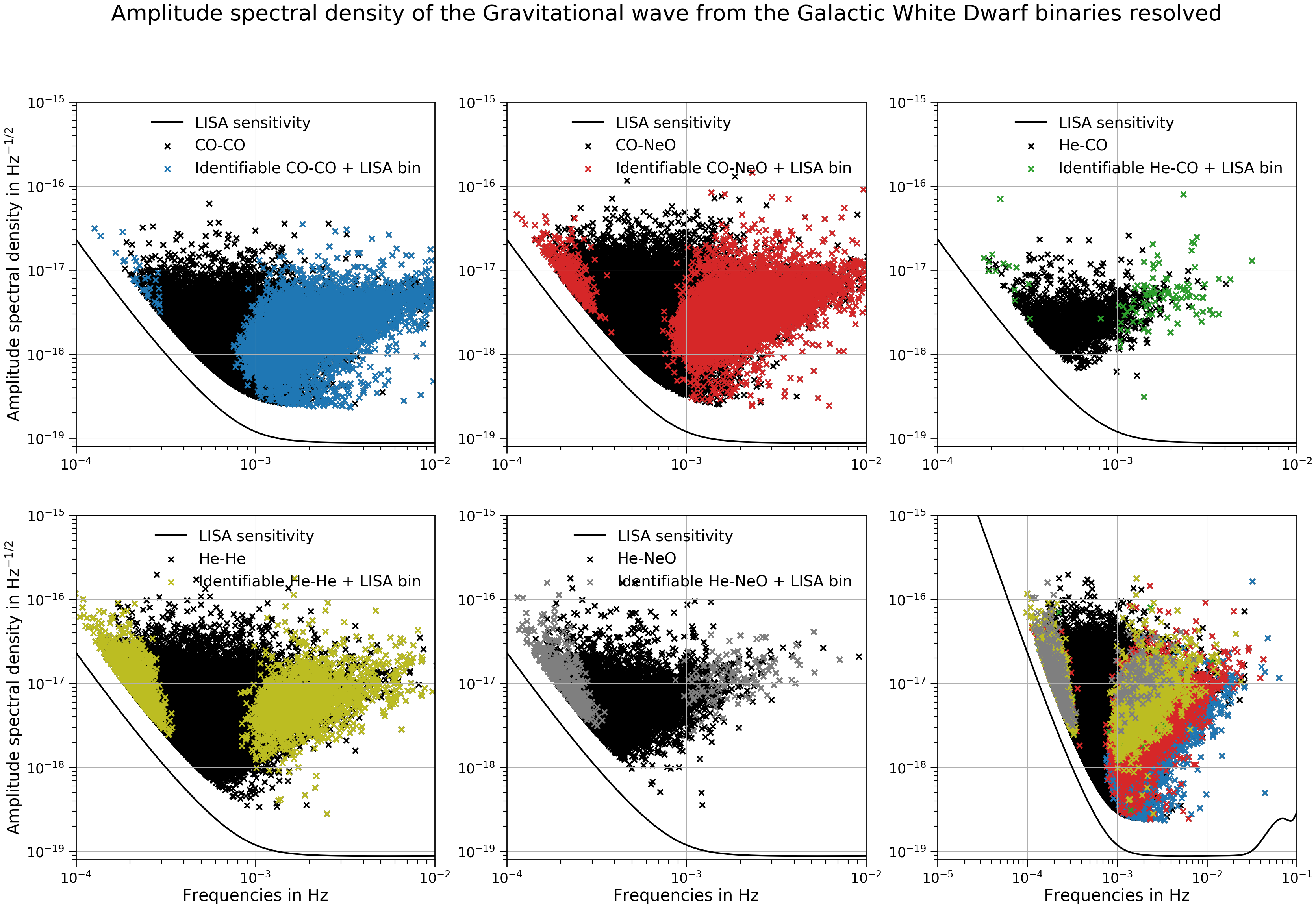}
    \caption{The GW amplitude spectral density for the different core compositions from the resolved Galactic DWD for four years of science time duration with the \textit{LISA} strain sensitivity $S_n(f)$.}
    \label{fig:ASDresol}
\end{figure*}

Fig.~\ref{fig:ASDresol} shows the GW amplitude spectral density for the different cores of DWD. We have evidence of the domination of the type He-CO and He-CO for the resolved binaries. The distributions for all binary types from the catalog, and the resolved binaries are presented in five plots. The black line corresponds to the \textit{LISA} strain sensitivity $S_n(f)$. The figure at the bottom right is the total distribution of binaries for the different core compositions. We cannot estimate the distribution of the unresolved binaries with the help of the distribution type of the resolved DWD. The gaps (black scatter) seen in the plots below $10^{-3}$ Hz correspond to the large number of sources close in terms of frequency. Indeed, this part of the spectrum comprises a large number of sources, that to be resolvable, must have a frequency difference greater than the frequency resolution of \textit{LISA}.

\subsection{Galactic confusion noise}

In Fig.~\ref{fig:ASD} the sensitivity curve can be further updated with the contribution from the Galactic confusion noise $S_c(f)$ \citep{2017JPhCS.840a2024C}, which corresponds to the unresolved binaries of the galactic population. This has been modeled with the catalog from \cite{2005MNRAS.356..753N} as a kind of broken power law. This model depends on the measurement duration, and for a duration of 4 years the model gives $\alpha = 0.138$, $\beta = -221$, $\kappa = 521$, $\gamma=1680$ and $f_k = 0.00113$:
\begin{equation}
    S_c(f) = A f^{-7/3} e^{-f^{\alpha} + \beta f \sin(\kappa f)}\left[1 + \tanh{\left(\gamma \left( f_k -f \right)\right)} \right].
\end{equation}
$A$ is the amplitude of the Galactic confusion noise in the low-frequency limit from the power spectrum of a quasi-circular binary population. This noise can be seen as adding further noise to the \textit{LISA} sensitivity.

In our calculation we introduce for each binary the amplitude gap from the other binaries in the local frequency band of the binary considered. We generate a catalog of resolved binaries; see Fig.~\ref{fig:ASDresol}. We note that the distribution of resolved binaries depends on the catalog used and the number of sources considered. 
Our study here is firstly an estimation of the ability to observe a cosmologically produced SGWB in the presence of a galactic foreground.
These estimates of resolved and unresolved binaries depend on the  catalog, but we find no evidence of any significant influence of the galactic foreground with the presence or not of resolved binaries in the foreground, or how they are defined (see Fig.~\ref{fig:PSDtot}).

\section{Calculation of the Waveform}\label{sec:waveform}

In this section we present the calculation of the waveform of the Galactic foreground.
 \subsection{Amplitude of the waveform}
\begin{figure}
    \centering
    \includegraphics[width=70mm]{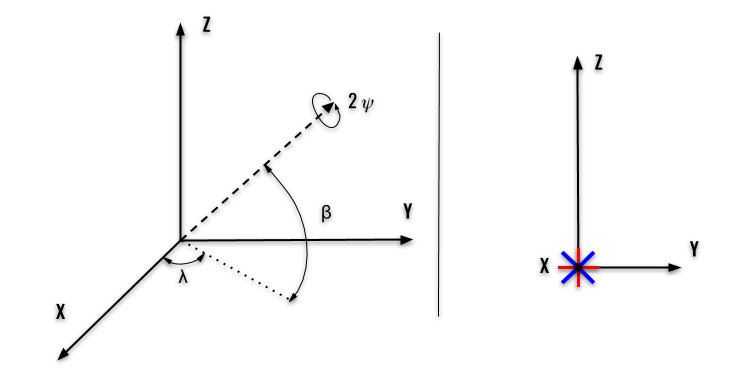}
    \caption{Generic plane waveform reference for a binary localised at ecliptic latitude $\beta$ and longitude $\lambda$. The \textit{LISA} constellation is in the center of the coordinate system. We use the conventional definition from \citet{PhysRevD.70.022003} for the GW polarisations: \textcolor{red}{$\bm{+}$} (in red) and  \textcolor{blue}{$\bm{\times}$} (in blue). The coordinate $\Psi$ is the rotation around the direction of wave propagation. }
    \label{fig:canvas}
\end{figure}

The GW strain $h(t)$ is given by the polarisation decomposition of the waveform,
 
\begin{equation}\label{eq:h(t)}
    \centering
    h(t) = h_+(t)\textbf{e}_+ + h_{\times}(t)\textbf{e}_{\times}
\end{equation}
where the two polarisation tensors $\textbf{e}_+$ and $\textbf{e}_{\times}$ given by:
\begin{equation}
   \textbf{e}_+ = E \begin{pmatrix}
    1&0&0\\
    0&-1&0\\
    0&0&0\\
\end{pmatrix} E^T   \hspace{.6cm}  \textbf{e}_{\times} = E \begin{pmatrix}
    0&1&0\\
    1&0&0\\
    0&0&0\\
\end{pmatrix} E^T
\end{equation}
with the polarisation coordinate matrix $E$; $\beta, \lambda$ are the ecliptic latitude and longitude, while $\psi$ is the rotation around the direction  of gravitational-wave propagation (see Fig~\ref{fig:canvas}),

\begin{equation}\label{eq:E}
E =  \begin{pmatrix}
    \lambda_s \psi_c - \lambda_c \beta_s \psi_s&- \lambda_s \psi_s - \lambda_c \beta_s \psi_c& - \lambda_c\beta_s\\
    -\lambda_c \psi_c - \lambda_s \beta_s \psi_s& \lambda_c \psi_s - \lambda_s \beta_s\psi_c&-\lambda_s\beta_c\\
    \beta_c\psi_s &\beta_c\psi_c&-\beta_s\\
\end{pmatrix}
\end{equation}
 The $s$ and $c$ subscripts refer to the sinus and cosinus operators, respectively.  

In Fig.~\ref{fig:canvas} we define the elliptical coordinates. For a DWD, the GW is a plane wave, quasi-monochromatic (there is a frequency drift over time). The frequency changes slightly for each orbit of the binary, the reason for this is energy loss from the emission of GW. For a DWD the drift is very small. The binary can be just considered as essentially monochromatic, and the polarisations given by:
\begin{equation}\label{eq:hA(t)}
\begin{pmatrix}
h_+(t) \\
h_{\times}(t) \\
\end{pmatrix} = \begin{pmatrix}
A_+(t) \cos(2 \pi ft +\Dot{f}t^2+\phi_0) \\
A_{\times}(t) \sin(2 \pi ft +\dot{f}t^2+\phi_0) \\
\end{pmatrix}
\end{equation}
with $\phi_0$ an initial phase. In the calculation we have a uniform distribution between 0 and 2$\pi$. The parameter $\dot{f}$ characterizes the frequency change from the loss of orbital energy.
To calculate the response of the detector arms, $\mathcal{H}_+(t)$ and $\mathcal{H}_{\times}(t)$, we need to calculate the one arm detector tensor $\textbf{D}$:
\begin{equation}
 \textbf{D} = \frac{1}{2} u\otimes u - v\otimes v
\end{equation}
where $u = \begin{pmatrix} 
1/2\\
0\\
\sqrt{3}/2\\
\end{pmatrix}$ and $v = \begin{pmatrix} 
-1/2\\
\sqrt{3}/2\\
0\\
\end{pmatrix}$.
Finally we have:
\begin{equation}
\begin{split}
&\mathcal{H}_+(t) = A_+(t) \cos(2 \pi ft +\phi_0)\textbf{e}_+:\textbf{D} \\
&\mathcal{H}_{\times}(t) = A_{\times}(t) \cos(2 \pi ft +\phi_0)\textbf{e}_{\times}:\textbf{D}
\end{split}
\end{equation}
where $\mathcal{H}_A(t) = h_A(t)\textbf{e}_A:\textbf{D}$ and A the two polarisations  $A=+,\times$.  $\mathcal{H}_A$ are the two polarisations in the detector basis.  This calculation of the galactic foreground is also presented in \citet{PhysRevD.76.083006}.

\subsection{Detector response function}

\begin{figure}
    \centering
    \includegraphics[width=70mm]{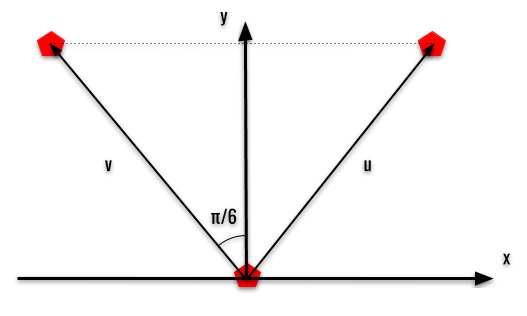}
    \caption{$(\hat{u}, \hat{v})$ coordinate system where the \textit{LISA} constellation is the red hexagons.}
    \label{fig:uvcoor}
\end{figure}

The detector response functions, $F_+$ and $F_{\times}$, for the location of the source at $(\theta,\phi)$ at the time $t$ in the basis vector $\hat{u}, \hat{v}$ are given by (see Fig.~\ref{fig:uvcoor}):
\begin{equation}
F_+ = -\frac{\sqrt{3}}{4}(1 + \cos(\theta))^2\sin(2\phi)
\end{equation}
\begin{equation}
F_{\times} = -\frac{\sqrt{3}}{2}\cos(\theta)\cos(2\phi)
\end{equation}
with $\hat{u}.\hat{z} = \sin(\phi + \pi/6)\sin\theta$ and $\hat{v}.\hat{z} = \sin(\phi - \pi/6)\sin\theta$, see \citet{Cornish_2001}.
The orbit of \textit{LISA} is one year around the sun, and also one year in revolution about itself (see Fig.~\ref{fig:LISA_constellation}). We need to consider the constellation orientation effects because \textit{LISA} will not see the sky uniformly.

\subsection{Signal of the DWD foreground measure by \textit{LISA}}\label{sc:signal}

\begin{figure*}
    \centering
    \includegraphics[height= 8cm]{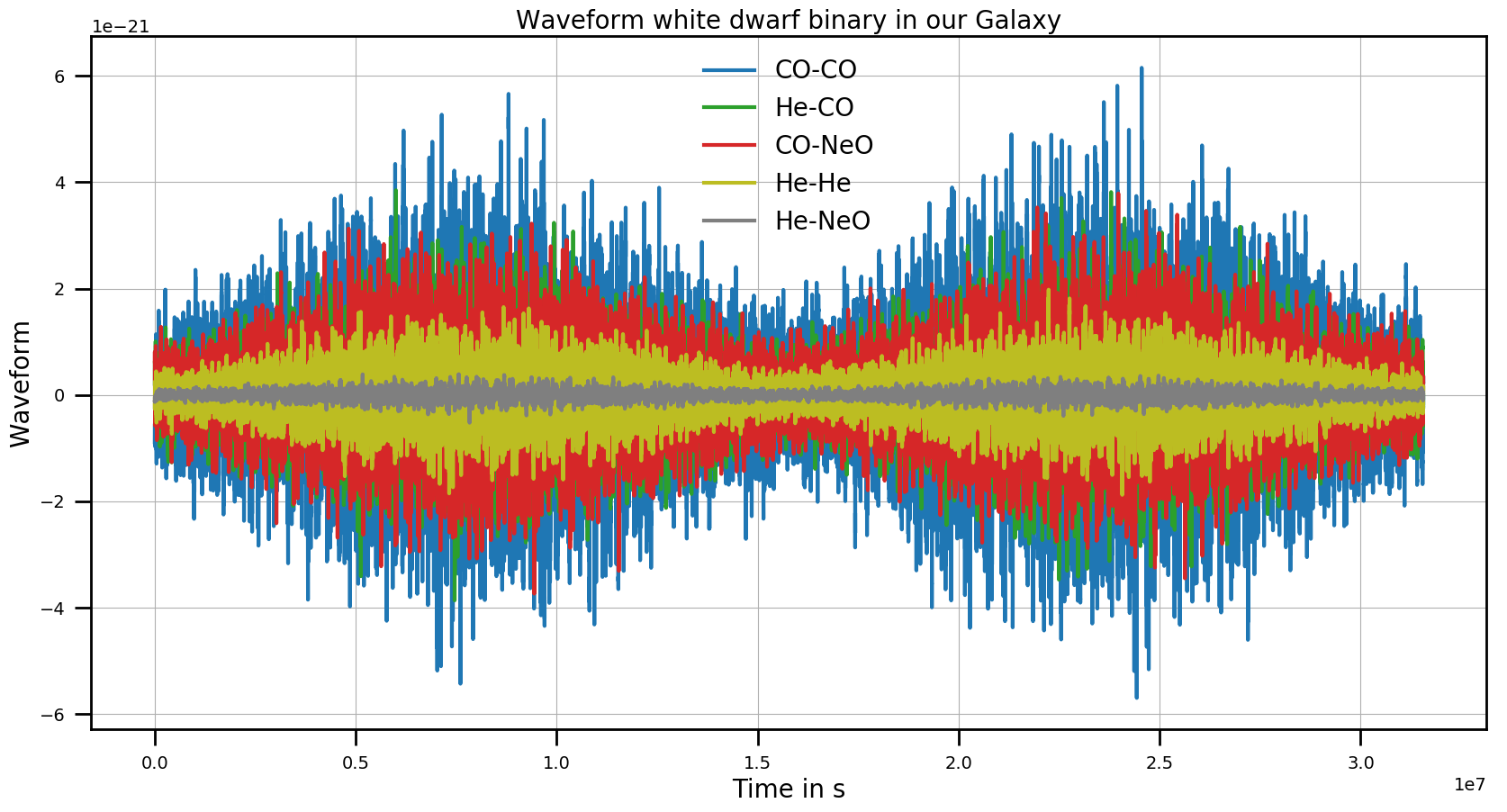}
    \caption{GW signals from the DWD modulation to be seen by \textit{LISA}. The modulation is from the evolution of the orientation of the \textit{LISA} constellation. The waveform $s(t)$ is the sum of the two polarisations $A = [\times, +]$ weighted by the respective detector response function $F^A(\Omega,f,t)$, such as $s(t) = \sum_N F^{\times}(\Omega,f,t)h_{\times}(f,t) + F^{+}(\Omega,f,t)h_{+}(f,t)$.}
    \label{fig:waveformtype}
\end{figure*}

\begin{figure*}
    \centering
    \includegraphics[height= 8cm]{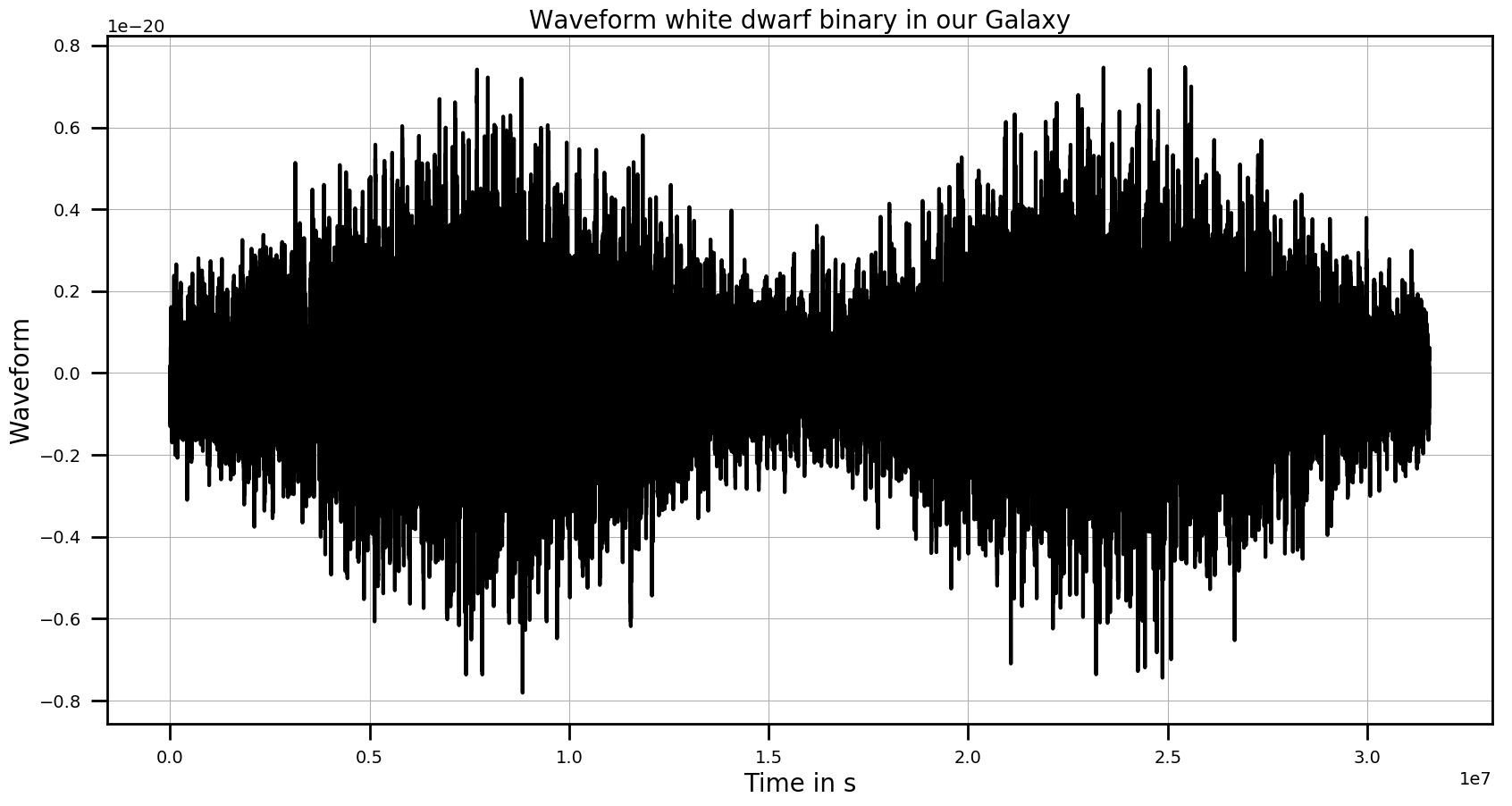}
    \caption{The total waveform, namely the sum of the five waveforms seen in Fig.~\ref{fig:waveformtype}, for the population from \citet{10.1093/mnras/stz2834} to be seen by \textit{LISA}.}
    \label{fig:totalwaveform}
\end{figure*}

We can build the total signal of the DWD foreground measured by \textit{LISA}; this is the sum of the waveforms for each DWD,
\begin{equation}
    \begin{split}
s(t) =& \sum_{i=1}^{N} \sum_{A= +,\times } h_{A,i}(f_{orb,i},M1_i,M2_i,X_i,Y_i,Z_i,t) \\ 
& \times F_A(\theta,\phi,t) \textbf{D}(\theta,\phi,f)_A:\textbf{e}_{A}
    \end{split}
\end{equation}
with $F_A$  the beam pattern function for the polarisations $A = +,\times$, $\textbf{h}_{A,i} = h_{A,i} \textbf{e}_{A}$ the tensor of the amplitude of the GW, and $\textbf{D}$ the one-arm detector tensor and $\textbf{h}_{A,i}$ the dimensionless GW amplitude of the binary $i$ (see Eqs.~\ref{eq:h(t)}~and~\ref{eq:hA(t)}).
Fig.~\ref{fig:waveformtype} shows the gravitational waveforms of the five populations of DWDs from \citet{10.1093/mnras/stz2834}. The waveform from CO-CO has the largest amplitude. The sum of the five populations becomes the waveform to be seen by \textit{LISA} (see Fig.~\ref{fig:totalwaveform}). The modulation of the DWD waveform is an orbital effect. In fact, when the \textit{LISA} constellation points toward the center of our Galaxy, the waveform amplitude will attain a maximum. Because of the symmetry of the plane passing through $\hat{u}$ and $\hat{v}$ (Fig.~\ref{fig:uvcoor}), this will happen twice per year. 

\section{Spectral Separation and study of the waveform}\label{sc:SpecSepa}

In this section we describe the calculation of the spectral energy density of the modulated DWDs, $\Omega_{GW,DWD}$, that were introduced in  Sec.~\ref{sec:waveform}. Fig.~\ref{fig:totalwaveform} is the modulated foreground for \textit{LISA}, with the DWDs from \citet{10.1093/mnras/stz2834}. The simulated DWD population resembles the DWD population of the Milky Way.

\subsection{Energy and Power Spectral Density} \label{sc:ESD}

Given the power spectral density (PSD), we can compute the energy spectral density of the Galactic foreground that \textit{LISA} will observe: 

\begin{equation}\label{eq:OmtoPSD}
    \Omega_{GW,DWD}(f) = \frac{4 \pi^2}{3 H_0^2}f^3 \frac{PSD(f)}{\mathcal{R}(f)} ~ ,
\end{equation}
with $H_0$ the Hubble-Lemaître constant ($H_0 \simeq 2.175 \ 10^{-18}  \ \text{Hz}$), $PSD(f)$ the power spectral density of the waveform of the Galactic foreground, and $\mathcal{R}(f)$ the \textit{LISA} response function. We use the periodogram to estimate the PSD. For a waveform $s(t)$ the periodogram is $I_n(f_k) = |\tilde{s}(f_k)|^2$, where $\tilde{s}(f_k)=\frac{1}{\sqrt{T}}\sum_{i=1}^T s(t) e^{-itf_k}$ at Fourier frequencies $f_k= 2\pi k/T,\; k=0,\ldots, N=\frac{T}{2}-1$ and $T$ the time duration of the signal for the different waveforms  (total binaries, resolved or unresolved binaries).  $\mathcal{R}(f)$ is the detector polarisation and sky averaged response function, which can be approximated by \citep{Robson_2019}:
\begin{equation}\label{eq:R}
    \mathcal{R}(f) \simeq \frac{3}{10} \frac{\left(\frac{2\pi f L}{c}\right)^2}{1+0.6 \left(\frac{2\pi f L}{c}\right)^2} ~ .
\end{equation}

\begin{figure*}
    \centering
    \includegraphics[height= 8cm]{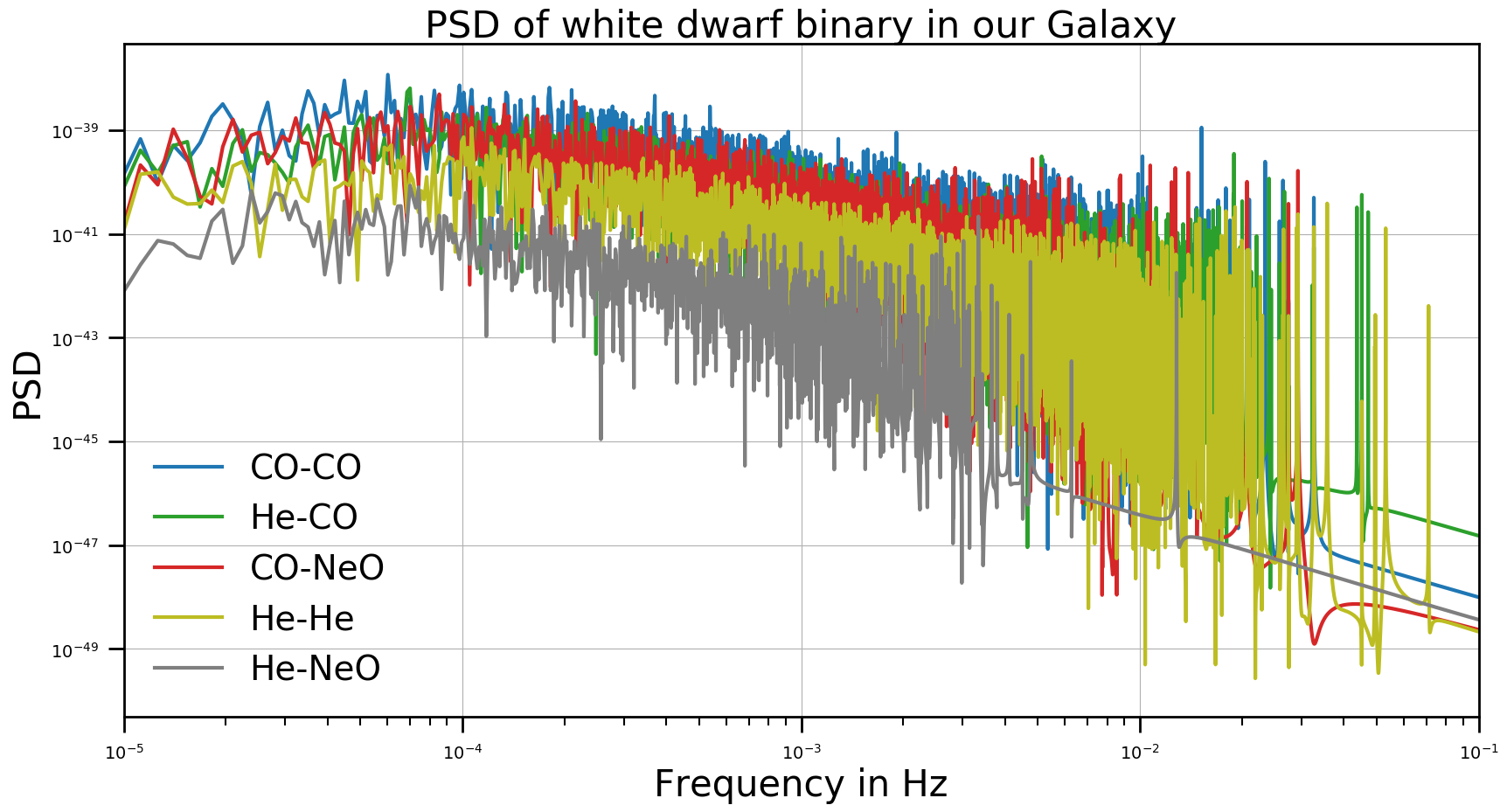}
    \caption{PSD of the GW signal for the WD binary modulation seen by \textit{LISA}, (see Fig.~\ref{fig:waveformtype}). The set of binaries in blue, green, red, yellow and grey are respectively the different cores CO-CO, He-CO, CO-NeO, He-He and He-NeO. This is the direct periodogram of Fig.~\ref{fig:waveformtype}.}
    \label{fig:PSD}
\end{figure*}

\begin{figure*}
    \centering
    \includegraphics[height= 8cm]{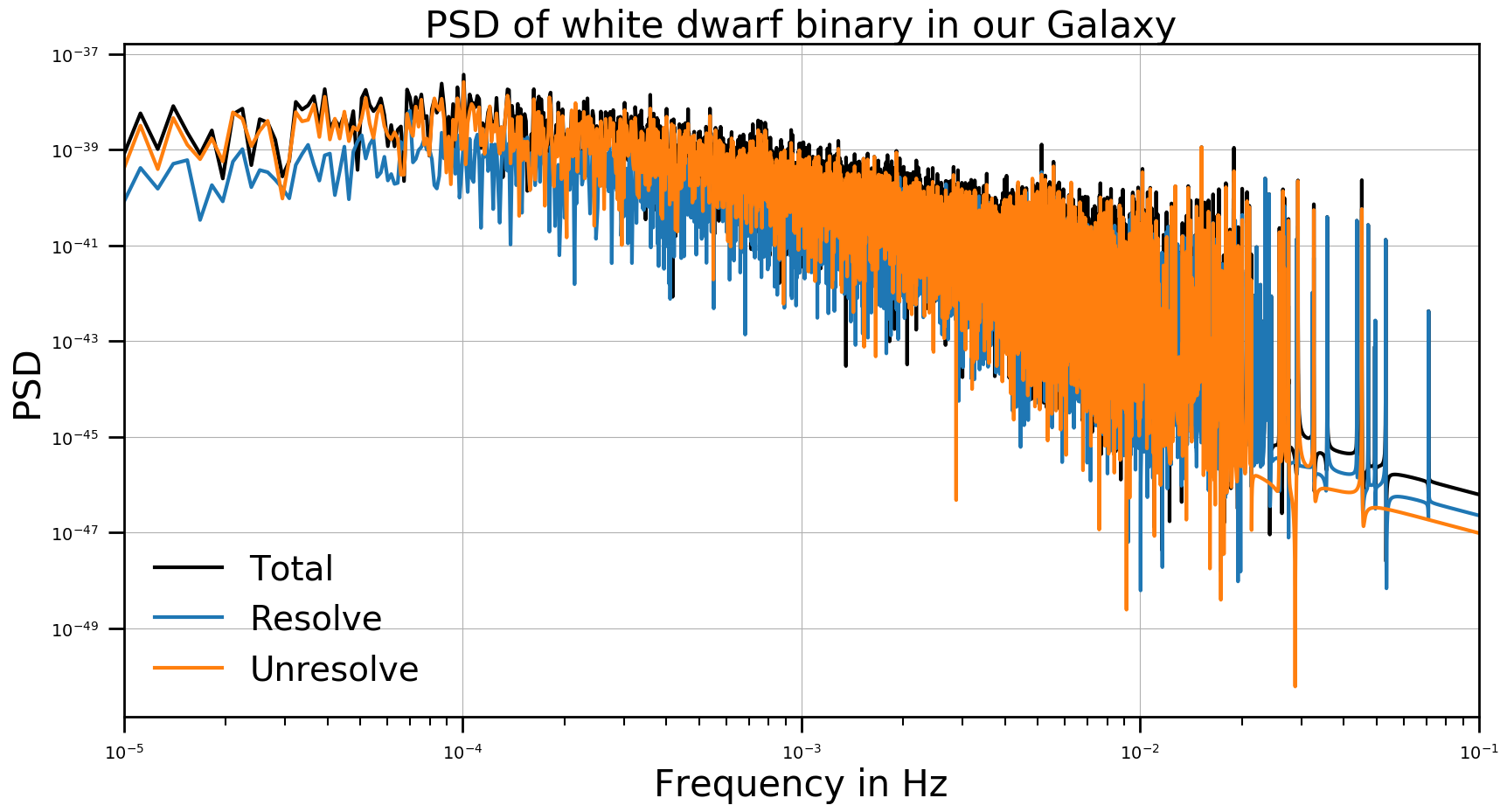}
    \caption{PSD of the GW signal for the WD binary modulation seen by \textit{LISA} (see Fig.~\ref{fig:waveformtype}). The set of binaries is in black, while the resolved and unresolved are respectively in blue and orange. These are the same as the Fig.~\ref{fig:ASD}.}
    \label{fig:PSDtot}
\end{figure*}

 The goal is to address the orbital motion of the \textit{LISA} constellation \citep{2003PhRvD..67b9905C}. 
 We can calculate this quantity with the mean square antenna pattern~\citep{Cornish_2001}, $\mathcal{R}(f) = \frac{1}{4\pi} \int \sum_A \mathbf{D}(\hat{\Omega},f):e^A(f) \mathrm{d}\hat{\Omega}$ \citep{2003PhRvD..67b9905C}.
Fig.~\ref{fig:PSD} gives the PSD for different types of WD cores, while Fig.~\ref{fig:PSDtot} presents the total PSD, plus the PSDs from resolvable and unresolvable DWDs.  
Fig.~\ref{fig:PSDtot} shows that the PSD of the total waveform is not purely a power law. According to \citet[Fig.~4]{2014PhRvD..89b2001A} and \citet[Fig.~1]{2020ApJ...901....4B}, we have fewer binaries at higher frequencies.

\subsection{Comparison of the energy spectral density from different catalogs}\label{sc:PSDs}
 
\begin{figure*}
    \centering
    \includegraphics[height= 8cm]{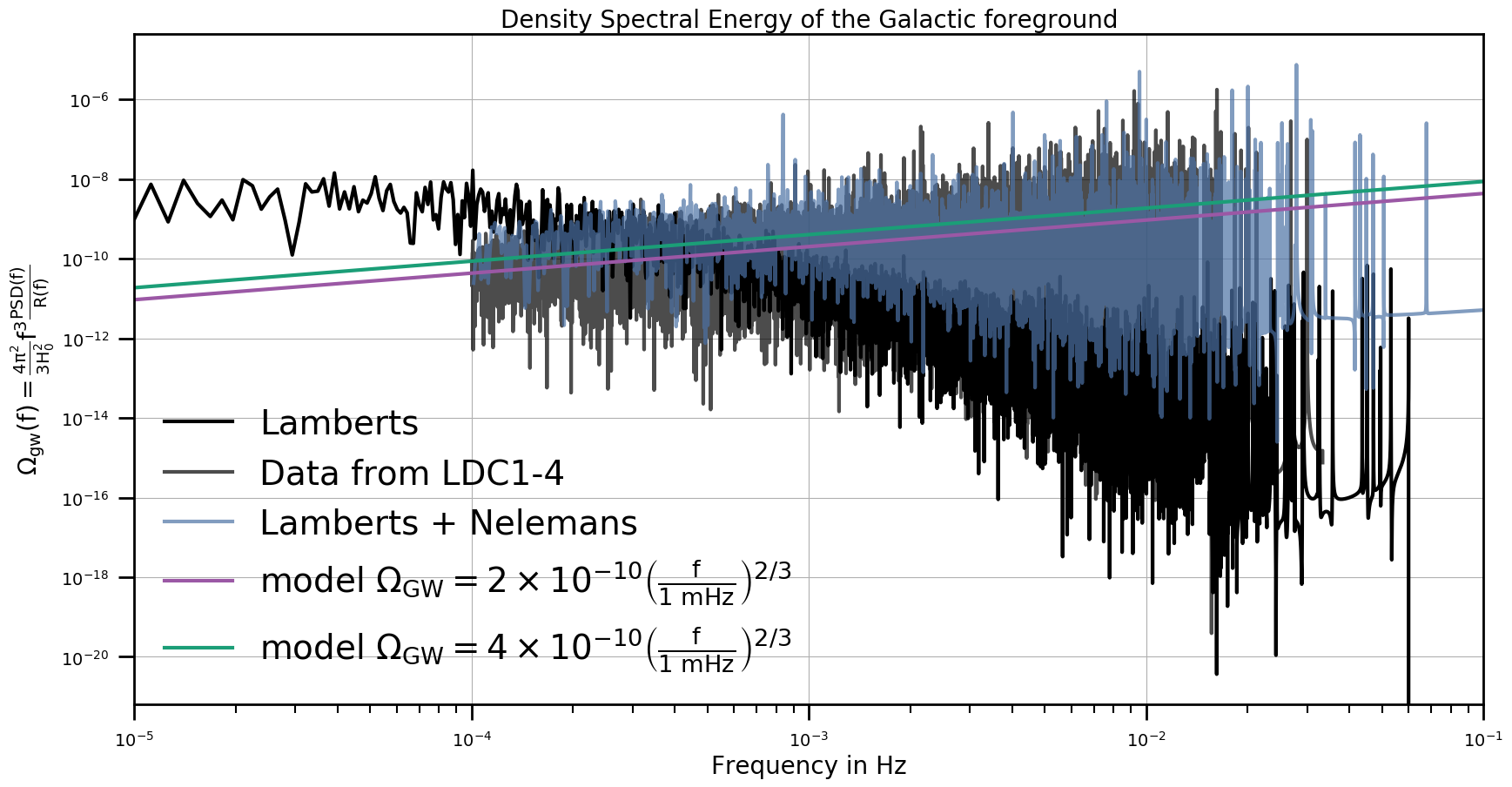}
    \caption{Normalised energy spectral density of the Galactic foreground $\Omega_{GW,DWD}$ for different population models of DWD in the Milky Way. The black line is the \citeauthor{10.1093/mnras/stz2834} population, the gray line corresponds to the population of LDC (LDC1-4); this population can be fit by a power law, see the purple line ($\Omega_{GW}(f) = 2 \times 10^{-10}\left(\frac{f}{1 \times 10^{-3} \ \text{Hz}}\right)^{2/3}$). In blue, labeled by Lamberts + Nelemans, is a population generated with binaries where the galactic spatial positions are given by \citeauthor{Nelemans_2001}, and the other parameters from \citeauthor{10.1093/mnras/stz2834}. This "Lamberts + Nelemans" population can be fit by a power law  $\Omega_{GW}(f) = 4 \times 10^{-10}\left(\frac{f}{1 \times 10^{-3} \ \text{Hz}}\right)^{2/3}$ in the frequency band $[1 \times 10^{-4} \ \text{Hz}, 1 \times 10^{-3} \ \text{Hz} ]$. }
    \label{fig:omegapops}
\end{figure*}

 In this section we derive the normalised energy spectral density of the Galactic foreground $\Omega_{GW,DWD}$ for different population models. We start with the population from the catalog \textit{LDC 1-4}. This population was simulated with the parameters from \citet{Nelemans_2001}. In Fig.~\ref{fig:omegapops} we show that at low frequencies the energy spectral density of the population from the \textit{LISA} Data challenge can be fit by a power law $\Omega_{GW,\textit{LDC 1-4}}(f) \simeq \Omega_{\textit{LDC 1-4}}\left( \frac{f}{f_{ref}}\right)^{\alpha}$, with  slope $\alpha=2/3$. This is the slope expected for a SGWB from binary systems (see grey line in Fig.~\ref{fig:omegapops}). The black line is the energy spectral density of the Galactic foreground for the population of \citet{10.1093/mnras/stz2834}. We have evidence that the power law can be fit at low frequencies (between $1 \times 10^{-5}$ and $1 \times 10^{-4}$ Hz) with the same slope $\alpha=2/3$, but at higher frequencies this power law breaks down. In order to understand whether the DWD spatial distribution model is responsible for this difference, we can use the population of \citet{10.1093/mnras/stz2834} but we use the DWD spatial distribution from \citet{Nelemans_2001}. This combination results in the blue line plot in Fig.~\ref{fig:omegapops}, labeled Lamberts + Nelemans. As an example we also display a purple line representing $\Omega_{GW}(f) = 2 \times 10^{-10}\left(\frac{f}{1 \times 10^{-3} \ \text{Hz}}\right)^{2/3}$ which lies over the LDC1-4 galactic foreground; this displays that the LDC1-4 galactic foreground can be approximate by a power law.
 
  The break down of the power law at high frequencies is due to the spatial distribution of the DWDs. Indeed, the blue line (\citet{10.1093/mnras/stz2834} + \citet{Nelemans_2001}) can be represented by a power law with a slope in $2/3$,  represented in Fig.~\ref{fig:omegapops} as the green line of $\Omega_{GW}(f) = 4 \times 10^{-10}\left(\frac{f}{1 \times 10^{-3} \ \text{Hz}}\right)^{2/3}$ in the frequency band $[1 \times 10^{-4} \ \text{Hz}, 1 \times 10^{-3} \ \text{Hz} ]$.

 The catalog of \citet{10.1093/mnras/stz2834} for the Milky Way DWD distribution cannot be represented as a power law over the entire \textit{LISA} spectral band. 
 However, we modify the function to better fit the scarcity of power at high frequencies, Eq.~\ref{eq:newmodel}, as a broken power-law, namely:
\begin{equation}\label{eq:newmodel}
    \Omega_{DWD}(A_1,\alpha_1, A_2, \alpha_2;f)= \frac{A_1 \left(\frac{f}{f_{ref}}\right)^{\alpha_1}}{1 + A_2 \left(\frac{f}{f_{ref}}\right)^{\alpha_2}}.
\end{equation}
In the remainder of this document we use $f_{ref} = \frac{2\pi L}{c} \simeq 0.019 \ \text{Hz}$.
For $1 \ll A_2 \left(\frac{f}{f_{ref}}\right)^{\alpha_2}$ (low frequencies) this yields:
\begin{equation}
     \Omega_{DWD}(f) \simeq \frac{A_1}{A_2} \left(\frac{f}{f_{ref}}\right)^{\alpha_1-\alpha_2} 
\end{equation}
i.e.\ the energy spectral density at low frequencies can be approximated by a power law function; for a DWD foreground the slope $\alpha = \alpha_1-\alpha_2$ has to be $\frac{2}{3}$.
For $1 \gg A_2 \left(\frac{f}{f_{ref}}\right)^{\alpha_2}$ (high frequencies):
\begin{equation}
     \Omega_{DWD}(f) \simeq A_1 \left(\frac{f}{f_{ref}}\right)^{\alpha_1} 
\end{equation}
i.e.\ for high frequencies the energy spectral density can also be approximated by a power law function but with different parameters. 
The differential $d\Omega(f)$ is therefore:
\begin{equation}\label{eq:domega}
\begin{split}
    d\Omega(f) &= \Omega \Bigg[ \frac{dA_1}{A_1} + \ln\left(\frac{f}{f_{ref}}\right)d\alpha_1 + \frac{dA_2\left(\frac{f}{f_{ref}}\right)^{\alpha_2}}{1 + A_2 \left(\frac{f}{f_{ref}}\right)^{\alpha_2}} \\
    &+  \frac{d\alpha_2A_2 \ln\left(\frac{f}{f_{ref}}\right)\left(\frac{f}{f_{ref}}\right)^{\alpha_2}}{1 + A_2 \left(\frac{f}{f_{ref}}\right)^{\alpha_2}}\Bigg].
\end{split}
\end{equation}
To estimate the four parameters  $\Omega(A_1,\alpha_1, A_2, \alpha_2;f)$ of the model, we use an adaptive MCMC algorithm.

\subsection{Stochastic Gravitational Wave Background}\label{sc:SGWB}

In the frequency band $[f,\ f+df]$, the normalised energy spectral density of an isotropic SGWB, $\Omega_{GW}(f)$, can be modeled as a frequency variation of the energy density of the GW $\rho_{GW}$. The energy spectral density is a function of the differential variation over the frequency of the energy density $\rho_{GW}$~\citep{PhysRevD.46.5250,doi:10.1146/annurev.nucl.54.070103.181251,Christensen_2018}. The distribution of the energy density over the frequency domain can be expressed as 
\begin{equation}
\label{eq:Omega}
        \begin{split}
        \Omega_{GW}(f) &= \frac{f}{\rho_c} \frac{\mathrm{d} \rho_{GW}}{\mathrm{d} \ln(f)}\\
        &= \sum_k \Omega_{GW,k}(f).
        \end{split}
\end{equation}
where $k$ refers to a specific SGWB and the critical density of the universe is $\rho_c = \frac{3 H_0^2c^2}{8\pi G}$. In this paper we approximate the SGWB energy spectral density as a sum of power laws. We also assume that the astrophysical (binary black hole produced) and cosmological backgrounds are isotropic. 
We have $\Omega_{GW} \simeq \sum_k A_{k}\left( \frac{f}{f_{ref}}\right)^{\alpha_k}$, where the energy spectral density amplitude of the component $k$ (representing the different SGWBs) is $A_k$, the respective slopes are $\alpha_k$, and $f_{ref}$ is some reference frequency.

The energy spectral density of the the cosmological background  should have a slope  $ \alpha \approx 0$. This is a good approximation for scale invariant processes and also for standard inflation, but certainly false for cosmic strings and turbulence. However, for our study here, we will model the cosmologically produced SGWB energy density with $ \alpha = 0 $. In addition, for a second isotropic SGWB, the compact binary product astrophysical background, we use $ \alpha = \frac{2}{3} $. According to \citet{Farmer:2003pa}, the slope is $ \alpha = \frac{2}{3} $ for quasi-circular binaries evolving purely under emission of GW. Eccentricity and environmental effects can alter the slope and gravitation frequencies of the binary. We also note the limitations of our cosmological power law model as phase transitions in the early universe need to be approximated with two-part power laws, with a traction between the rising and falling power law component at some particular frequency peak. However, as a starting point we use two isotopic backgrounds, each described by a simple power law. Since the two backgrounds are superimposed, the task is to simultaneously extract the astrophysical and cosmological components, that is, to simultaneously estimate the astrophysical and cosmological contributions to the energy spectral density of the SGWB. In addition to the two isotropic sources we will also consider the Galactic foreground, but as a broken power law. 

For the purpose of this spectral separation study we will define the astrophysical background of GWs coming from the unresolved compact objects with the estimates of \cite{2019ApJ...871...97C}, $\Omega_{GW,astro}(f) = 4.4  \times 10^{-12}\left( \frac{f}{3 \times 10^{-3} \ \text{Hz}} \right)^{2/3}$. 

The amplitude of the cosmological background is a free parameter, and our goal is to determine how well \textit{LISA} will do in providing an estimate for this parameter given the astrophysical background and Galactic foreground, plus \textit{LISA} detector noise.

The spectral separability study of \citet{2020arXiv201105055B} was recently carried out in the context of \textit{LISA}. There it was shown that it is possible for \textit{LISA} to measure a cosmological background with an amplitude between $1\times 10^{-13}$ and $1\times10^{-12}$ in the presence of the astrophysical background and \textit{LISA} detector noise.  
In this present paper we include a galactic foreground, and we model the energy spectral density of  the SGWB by

\begin{equation}
\label{eq:sum_all_Omegas}
    \begin{split}
        \Omega_{GW}(f) &= \Omega_{GW,DWD}(f) + \Omega_{GW,astro}(f) + \Omega_{GW,cosmo}(f) \\
        &= \frac{A_1 \left(\frac{f}{f_{ref}}\right)^{\alpha_1}}{1 + A_2 \left(\frac{f}{f_{ref}}\right)^{\alpha_2}} + \Omega_{astro} \left(\frac{f}{f_{ref}}\right)^{\alpha_{astro}} \\
        &+ \Omega_{cosmo} \left(\frac{f}{f_{ref}}\right)^{\alpha_{cosmo}}.
    \end{split}
\end{equation}

\subsection{Adaptive Markov Chain Monte-Carlo} \label{sc:AMcMC}

\subsubsection{Markov Chain Monte-Carlo}

Bayesian inference is a method by which the probability distribution of various parameters are determined given the observation of events. It is based on Bayes' theorem (see Eq.~\ref{eq:bayes}). 
The goal of a Bayesian study is to derive the posterior distribution of the parameters after observing the data.
According to Bayes' theorem, this is proportional to the likelihood, i.e.\ the distribution of the observations $d$ given the unknown parameters $x$ of our model, and the prior distribution of the parameters. 
Bayes' theorem is given by:
\begin{equation}\label{eq:bayes}
    p(x|d) = \frac{p(d|x)p(x)}{p(d)}
\end{equation}
where $p(x)$ is the prior distribution, $p(x|d)$ is the posterior distribution, $p(d|x)$ is the likelihood, and $p(d)$ is the evidence.
 There are various sampling-based strategies for calculating the posterior distribution, so called MCMC methods \citep{1953JChPh..21.1087M,gilks1995markov}.

\subsubsection{Metropolis-Hasting sampler}

MCMC methods are based on the simulation of a Markov chain.
To simulate from a Markov chain, we use the
Metropolis-Hastings algorithm \citep{10.1093/biomet/57.1.97,gilks1995markov}. This is based on the rejection or acceptance of candidate parameters according to the likelihood ratio
between two neighboring candidates. Thus,  candidate parameters with higher values of the posterior distribution are favored but candidates with lower values are accepted with a certain probability given by the Metropolis-Hastings ratio below:\newline

\textbf{Metropolis-Hastings algorithm}
    \begin{itemize}
        \item initial point $x_0$ 
        \item at the i-th iteration:
            \begin{itemize}
                \item Generate candidate $x'$ from symmetric proposal density $g(x'|x_i)$ (e.g.\ Gaussian with mean $x_i$)
                \item Evaluation 
                    \begin{itemize}
                        \item likelihood of $x_i$ and $x'$, $p(d|x_i)$ and $p(d|x')$
                        \item prior of $x_i$ and $x'$, $p(x')$ and $p(x_i)$
                        \item ratio $\alpha = \frac{p(d|m(x'))}{p(d|m(x_i))} \frac{p(x')}{p(x_i)}$
                    \end{itemize}
                \item Accept/Reject
                \begin{itemize}
                    \item Generation of a uniform random number $u$ on $[0,1]$
                    \item if $u \leq \alpha$, accept the candidate : $x_{i+1} = x'$
                    \item if $u > \alpha$, recycle the previous value : $x_{i+1} = x_i$
                \end{itemize}
            \end{itemize}
    \end{itemize}
    
At the end of the algorithm, we have a certain acceptance rate. If this rate is too close to 0, it means that the Markov chain made frequent large moves into the tails of the posterior distribution which got rejected and therefore it mixed only very slowly; if the acceptance rate is too close to 1, the Markov chain made only small steps which had a high probability of getting accepted but it took a long time to traverse the entire parameter space. In either of these cases, the convergence towards the stationary distribution of the Markov chain will be slow. To control the acceptance rate we can introduce a step-size parameter; this is often the standard deviation of the jump proposal $g(x'|x_i)$. 
This step-size parameter can be modified `on the fly' while the algorithm is running to improve the exploration of the parameter space. Similarly, the proposal density should take correlations between the parameters into account to improve mixing. These can also be estimated `on the fly' based on the previous values of the Markov chain and convergence of such an adaptive MCMC is guaranteed as long as a diminishing adaptation condition is met \citep{doi:10.1198/jcgs.2009.06134}.
Post convergence of the MCMC, a histogram or kernel density estimate based on the samples of each parameter provides an estimate of its posterior density.  To summarize the posterior distribution, the sample mean of the Markov chain gives a consistent estimate of its expectation. Similarly the sample standard deviation provides an estimate for its standard error. It is important to check whether the posterior distribution is different from the prior because otherwise the data will not have provided any additional information about the unknown parameters beyond that of the prior distribution.

\subsubsection{Adaptive Markov Chain Monte-Carlo} \label{sec:adaptive}

We use the A-MCMC version from the \textit{Examples of Adaptive MCMC } \citep{doi:10.1198/jcgs.2009.06134}. For a $d$-dimensional MCMC we can perform the Metropolis-Hasting sampling with a proposal density $g_n(x)$ defined by:
\begin{equation}
    g_n(x)= (1-\beta)N(x,(2.28)^2 \Sigma_n / d ) + \beta N(x,(0.1)^2 I_d/d)
\end{equation}
with $\Sigma_n$ the current empirical estimate of the covariance matrix obtained from the samples of the Markov chain, $\beta = 0.25$ a constant, 
$d$ the number of parameters, $N$ the multi-normal distribution, and $I_d$ the identity matrix. We estimate the covariance matrix based on the last one hundred samples of the chains. 

\section{\textit{LISA} stochastic gravitational wave background fitting with Adaptive Markov chain Monte-Carlo}\label{sc:cov}
\label{sec:AET}

In this section we consider the \textit{LISA} null channel $T$, and the science channels $A$ and $E$. We assume that the observation of the noise in channel $T$ informs us of the noise in channels $A$ and $E$. We follow the formalism of \citet{PhysRevD.100.104055}. 
Channels $A, E$ and $T$ are derived from channels $X$, $Y$ and $Z$ \citep{Vallisneri_2012}, the unequal-arm Michelsons centered on the three spacecraft. 

We assume that the noise is uncorrelated between the channels $X,Y,Z$. Nominally, the channel $T$ does not contain a GW signal, but contains the uncorrelated \textit{LISA} noise.  
This assumption is not perfectly accurate, but for this analysis we will hold it to be true. 
The relations are:
\begin{equation} 
    \left\{
\begin{array}{l}
     A = \frac{1}{\sqrt{2}}(Z-X) \\
     E = \frac{1}{\sqrt{6}}(X-2Y+Z) \\
     T = \frac{1}{\sqrt{3}}(X+Y+Z).
\end{array}
\right.
\label{eq:AET} 
\end{equation}
We assume that channels $A$ and $E$ contain the same GW information and the same noise PSD.
Channel $T$ has a different noise PSD but shares parameters with the noise PSDs of channels $A$ and $E$. Thus, in the context of a study of spectral separation of the SGWB, the use of the $T$ channel allows us to simultaneously estimate the noise parameters for the \textit{LISA} channels $A$ and $E$ more accurately. Without the simultaneous estimation of the LISA noise parameters from the $T$ channel, it is possible to estimate the \textit{LISA} noise parameters, but this estimate becomes less accurate. As a result, the SGWB parameter estimate accuracy is degraded. It is in this sense that it is also important to use the $T$ channel.

This was also the motivation for the study of~\citet{2020arXiv201105055B}. 

We simulate the noise and SGWB in the frequency domain assuming
\begin{equation}\label{eq:modelPSDs}
\left\{
\begin{array}{l}
    PSD_A = S_A + N_A, \\
    PSD_E = S_E + N_E, \\
  PSD_T = N_T
\end{array}
\right.
\end{equation}
with \\
$S_A(f) = S_E(f) =\frac{3H_0^2}{4 \pi^2} \frac{ \sum_i \Omega_{GW,i}\mathcal{R}(f)}{f^3}$, $f_{ref}=25 \ \text{Hz}$, and \\
$\Omega_{GW} = \Omega_{GW,0}\left(\frac{f}{f_{ref}}\right)^{\alpha_0} + \Omega_{GW,2/3}\left(\frac{f}{f_{ref}}\right)^{\alpha_{2/3}} + \frac{A_1 \left(\frac{f}{f_{ref}}\right)^{\alpha_1}}{1 + A_2 \left(\frac{f}{f_{ref}}\right)^{\alpha_2}}$. \\
The noise components $N_A(f)=N_E(f)$ and $N_T(f)$ can be written as:
\begin{equation}
\left\{
\begin{array}{l}
    N_A = N_1 - N_2, \\
    N_T = N_1 + 2 N_2 
\end{array}
\right.
\end{equation}
with 
\begin{equation}
\left\{
\begin{array}{l}
    N_1(f) = \left(4 S_s(f) + 8\left( 1 + \cos^2\left(\frac{f}{f_{ref}}\right)\right) S_a(f)\right)|W(f)|^2 \\
    N_2(f) = -\left(2 S_s(f) + 8 S_a(f)\right)\cos\left(\frac{f}{f_{ref}}\right)|W(f)|^2
\end{array}
\right.
\end{equation}
$W(f) = 1 - e^{-\frac{2if}{f_{ref}}}$ and 
\begin{equation}
\left\{
\begin{array}{l}
    S_s(f) = N_{Pos} \\
    S_a(f) = \frac{N_{acc}}{(2 \pi f)^4}\left( 1 + \left(\frac{4 \times 10^{-4} \ \text{Hz}}{f} \right)^2 \right).
\end{array}
\right.
\end{equation}
The magnitude of the level of the \textit{LISA} noise budget is specified by the \href{https://atrium.in2p3.fr/nuxeo/nxdoc/default/f5a78d3e-9e19-47a5-aa11-51c81d370f5f/view_documents}{\textit{LISA} Science Requirement Document}  and \citet{2019arXiv190706482B}. To create the data for our example we use an acceleration noise of $N_{acc} = 1.44 \times 10^{-48} \ \text{s}^{-4}\text{Hz}^{-1}$ and the optical path-length fluctuation $N_{Pos} = 3.6 \times 10^{-41} \ \text{Hz}^{-1}$. 

We use this simplified model to generate the \textit{LISA} noise. We note that the  \textit{LISA} noise model for the channels $A$, $E$, and $T$ from TDI, as in the LDC \citep{2010PhRvD..82b2002A,LDC} matches the \textit{LISA} noise model of \citet{PhysRevD.100.104055}.
Once again, the goal of our study that we present in this paper here is to display the methods of spectral separability and determine the level of detectabilty of a cosmologically produced SGWB in the presence of an astrophysically produced SGWB, a Galactic foreground, and \textit{LISA} noise. We leave for future work an increase in complexity of the \textit{LISA} noise.

Our model contains  ten unknown parameters: $\theta = (N_{acc}, N_{Pos}, A_1, \alpha_1, A_2, \alpha_2, \Omega_{astro}, \alpha_{astro},\Omega_{cosmo}, \alpha_{cosmo})$. 
We  calculate the propagation of uncertainties for the power spectral densities with the partial derivative method. As such, we can estimate the induced uncertainty about the PSD resulting from estimation uncertainty of $\theta$, $dPSD = \sqrt{\sum_{\theta} \left(\frac{\partial PSD} {\partial \theta}\right)^2 d\theta^2}$. We then obtain for three SGWB components
$\Omega_{GW,astro}(f) = \Omega_{astro}\left(\frac{f}{f_{ref}}\right)^{\alpha_{astro}}$, $\Omega_{GW,cosmo}(f) = \Omega_{cosmo}\left(\frac{f}{f_{ref}}\right)^{\alpha_{cosmo}}$ and $\Omega_{DWD}(f)= \frac{A_1 \left(\frac{f}{f_{ref}}\right)^{\alpha_1}}{1 + A_2 \left(\frac{f}{f_{ref}}\right)^{\alpha_2}}$. 
\begin{equation}\label{eq:errorPSD}
\left \{
\begin{array}{l}
    \begin{split}
        dPSD_I &= \Big[N_I(0, dN_{acc},f)^2 + N_I(dN_{pos}, 0,f)^2 \\
        &+ S_I(\Omega_{astro}, \alpha_{astro},\Omega_{cosmo}, \alpha_{cosmo},f)^2 \Big(d\Omega_{cosmo}^2 \\
        &+ d\Omega_{astro}^2+ \ln\left(\frac{f}{f_{ref}}\right)^2\Big(\Omega_{astro}^2 d\alpha_{astro}^2  \\ 
        &+ \Omega_{cosmo}^2 d\alpha_{cosmo}^2\Big) \Big) + S_{DWD}^2(f) \Bigg( \left(\frac{dA_1}{A_1}\right)^2 + \ln\left(\frac{f}{f_{ref}}\right)^2  \\
        &\times d\alpha_1^2+ \left( A_2^2d\alpha_{2}^2 + dA_2^2\right)\left(\frac{\left(\frac{f}{f_{ref}}\right)^{\alpha_2}}{1 + A_2 \left(\frac{f}{f_{ref}}\right)^{\alpha_2}}\right)^2 \Bigg) \Big]^{1/2}\\
    \end{split}\\
    dPSD_T = \Big[N_T(0, dN_{acc},f)^2 + N_T(dN_{pos}, 0,f)^2\Big]^{1/2}
\end{array}
\right.
\end{equation}
with  $I = A,E$ and $[dN_{acc},$ $dN_{pos},$ $d\Omega_{astro},$ $d\alpha_{astro},$ $d\Omega_{cosmo},$ $d\alpha_{cosmo},$ $dA_1,$ $d\alpha_1,$ $dA_2,$ $d\alpha_2 ]$ are the standard deviations of the posterior distribution, assumed to be Gaussian.

To sample from their joint posterior distribution, we use the adaptive MCMC method  described in \ref{sc:AMcMC}. The sampled chains provide estimates for the standard error bands
for the power spectral density. With the MCMC chains we can calculate a histogram of $PSD_I(f)$ for each frequency. On each histogram we can then compute the $68\%$ credible band. This method is extracted from the BayesWave~\citep{Cornish:2014kda}, see Fig.~6 of the LIGO-Virgo data analysis guide paper \citet{2020CQGra..37e5002A}. The two methods give the same estimation for the errors.

For each pair $I,J = [A,E,T]$ we can calculate the covariance of $(\tilde{d}_I(f),\tilde{d}_J(f)$:
\begin{equation}
    <PSD_I(f),PSD_J(f)> = \mathcal{C}_{I,J}(\theta,f) 
\end{equation}

yielding the covariance matrix of $(\tilde{d}_A(f),\tilde{d}_E(f),\tilde{d}_T(f))$
(omitting the dependence of $f$ in the notation)
\begin{equation}
     \mathcal{C}(\theta,f) =
     \left(
     \begin{array}{ccc}
      S_A + N_A & 0 & 0  \\
      0 & S_E + N_E & 0 \\
      0 & 0 & N_T \\
     \end{array}
     \right)
\end{equation}
\begin{equation}
     \mathcal{C}^{-1}(\theta, f) = K
     \left(
     \begin{array}{ccc}
      (S_A + N_A)^{-1} & 0 & 0  \\
      0 &  (S_E + N_E)^{-1} & 0  \\
      0 & 0 &  N_T^{-1} \\
     \end{array}
     \right)
 \end{equation}
 and $K = det(\mathcal{C}) = \frac{1}{(S_A + N_A)(S_E + N_E)N_T}$.
 Using the definition of the Whittle likelihood as in~\cite{Romano2017},  the log-likelihood is given by:
\begin{equation} 
\begin{split}
     \mathcal{L}(\textbf{d}|\theta) & =  -\frac{1}{2} \sum_{k=0}^N \Bigg[ \sum_{I,J = [A,E,T]} \left( \sqrt{d_I(f)} \left(\mathcal{C}^{-1}\right)_{IJ} \sqrt{d_J(f)} \right) \\
     &+ \ln\left(2\pi K(f_k) \right) \Bigg] \\
    &=  -\frac{1}{2} \sum_{k=0}^N \Bigg[ \frac{d_A^2}{S_A+N_A}  + \frac{d_E^2}{S_E+N_E} + \frac{d_T^2}{N_T} \\
    &+  \ln\left(8\pi^3 (S_A+NA)(S_E+N_E)N_T \right) \Bigg] 
\end{split}
\end{equation}

As described in \citet{2020arXiv201105055B}, the inverse of the Fisher information matrix 
\begin{equation}\label{eq:fisher}
    \begin{split}
        F_{ab} &=   \frac{1}{2} \mathrm{Tr}\left(\mathcal{C}^{-1}\frac{\partial \mathcal{C}} {\partial \theta_a} \mathcal{C}^{-1} \frac{\partial \mathcal{C}}{\partial\theta_b}  \right) \\
        &= M \sum_{k=0}^{N} \Bigg[\frac{\frac{\partial (S_A+N_A)} {\partial \theta_a}\frac{\partial (S_A+N_A)} {\partial \theta_b}}{2(S_A+N_A)^2} \\
        &+ \frac{\frac{\partial (S_E+N_E)} {\partial \theta_a}\frac{\partial (S_E+N_E)} {\partial \theta_b}}{2(S_E+N_E)^2} + \frac{\frac{\partial N_T} {\partial \theta_a}\frac{\partial N_T} {\partial \theta_b}}{2N_T^2}\Bigg]
    \end{split}
\end{equation}
is the asymptotic covariance matrix of the parameters, $F^{-1}_{ab} = Cov(a,b)$.
The diagonal of this matrix gives the square of the standard error of the parameters, with $M=Df_b$ ($D$ is the time duration of the LISA mission and $f_b$ the highest frequency of interest in the LISA band). Thus, we can calculate the estimation uncertainty of a parameter in addition to using MCMC, and then show the consistency of the two sets or results.

\section{Measure of the amplitude of the DWD in the context of the orbital modulation}\label{sc:modulation}

In this section, we focus on the measurement of the GW signal amplitude coming from our Galaxy \citep{PhysRevD.71.122003}. In light of the waveform (see Fig.~\ref{fig:waveformtype}), we note that
if we want to estimate it correctly, we must account for the orbital motion of the \textit{LISA} constellation. The waveform is modulated in amplitude, and this modulation comes from the anisotropy of the Galactic foreground. The response of \textit{LISA} depends on its antenna pattern; the response is not homogeneous over the sky. At a particular time the \textit{LISA} response is maximal for sources localized on the orthogonal line passing through the center of the triangle which forms the \textit{LISA} constellation. This is like a "line of sight". The \textit{LISA} constellation is also in orbit around the sun and turns on itself. The line of sight does not always point in the same direction. If the GW background is isotropic, the detected amplitude is constant over time. 

We cut the waveform into small-time sections, 50 per year, and assume that the waveform is approximately constant within these sections. For each section, we calculate the signal standard deviation. This approximation works well and is consistent with the results of \citet{2014PhRvD..89b2001A}, and allows for a good fit of the orbital modulation.

\begin{figure*}
\centering
\includegraphics[height=9.0cm]{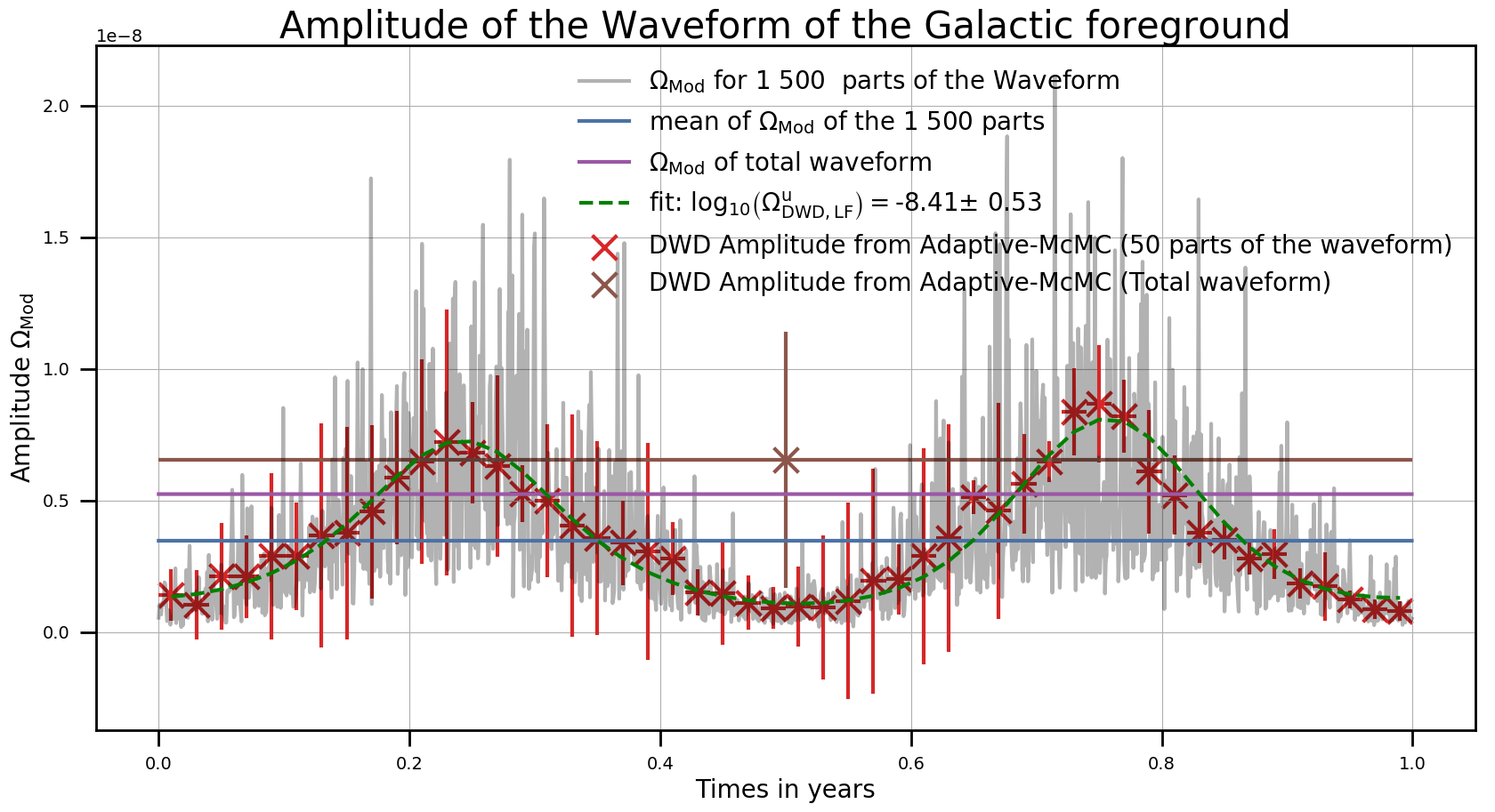}
\caption{Measurement of the orbital modulation of the DWD amplitude. In grey: 1500 estimations of the amplitude $\Omega_{Mod,i} = \frac{4\pi^2}{3H_0^2}\left(\frac{c}{2\pi L }\right)^2 A_i^2$, where $A_i$ is the amplitude of the characteristic strain of the section $h_i$ of the waveform. The blue line is the mean of the 1500 estimations of the amplitude. In purple, the amplitude $\Omega_{Mod}$ for the total waveform (1 year duration). This corresponds to the estimation of the amplitude of the energy spectral density of the DWD, but expressed as a time-series. In red scatter: 50 A-MCMC generated DWD amplitudes calculated from the estimates of the 8 parameters (BBH + WD + \textit{LISA} noise). In brown scatter, the estimation of the amplitude energy spectral density of the Galactic foreground at law frequency with the adaptive MCMC. We also fit the 50 A-MCMC with a least squares method to estimate the modulation from the \textit{LISA} antenna pattern and the 'real' amplitude of the Galactic foreground at low frequencies $\log_{10}\left(\Omega_{DWD,LF}^u\right) = -8.41 \pm 0.53$ (see Eq.~\ref{eq:fitMod}) for a reference frequency (see Eq.~\ref{eq:sum_all_Omegas}) of $3 \times 10^{-3} \ \text{Hz}*$(green dashed line).}
\label{fig:std_waveform}
\end{figure*}

It is then possible to calculate the amplitude of the spectral energy density, as by the methods of \citet{2013PhRvD..88l4032T}:
\begin{equation}
    \Omega_{Mod,i}=  \frac{4\pi^2}{3H_0} \left( \frac{c}{2 \pi L}\right)^2 A_i^2
\end{equation}
where $\Omega_{Mod,i}$ is the amplitude of the spectral energy density of the Galactic foreground at low frequencies for the section $i$ of the waveform; this is plotted over a year in Fig.~\ref{fig:std_waveform}, and the modulation from the \textit{LISA} orbit is apparent. Here $A_i$ is the amplitude of the characteristic strain for section $i$ if we consider the characteristic strain of the section $i$ as $h_{c,i} = A_{i}  \left(\frac{f 2 \pi L}{c}\right)^{\alpha}$, and the relation between the power spectral density $S_{h,i}(f)$ of the time series $h_i(t)$ and the spectral energy density of the section $i$, $\Omega_{GW,i}$ is $S_{h,i}(f) = \frac{3H_0^2}{4 \pi^2} \frac{\Omega_{GW,i}}{f^3}$, where $\Omega_{GW,i} = \Omega_{Mod,i}\left(\frac{f 2 \pi L}{c}\right)^{2\alpha +2}$. The goal is to estimate the level of the amplitude and compute the Galactic contribution to the sum of all the stochastic background and noise signals for \textit{LISA}. 

The grey curve in Fig.~\ref{fig:std_waveform} is the amplitude of the spectral energy density calculated with this method. We cut the year long time series into 1500 sections. We observe the amplitude modulation, indeed it is always maximum when \textit{LISA} points to the center of our Galaxy. The blue line corresponds to the mean of the 1500 estimates of the DWD amplitude. The purple line is the estimate of the DWD amplitude DWD 
for the total waveform, (1 year). 

As in  \citet[Figure~2]{2014PhRvD..89b2001A}, we measure the amplitude of the Galactic foreground for 50 sections of the year-long observation. In order to study the spectral separation of the stochastic background, we add the \textit{LISA} noise \citep{PhysRevD.100.104055} and an astrophysical background \citep{2019ApJ...871...97C} to the Galactic foreground and estimate the 8 parameters with an A-MCMC. 
The red scatterplot in Fig.~\ref{fig:std_waveform} is the estimate of the amplitude of the Galactic foreground in the low frequency approximation based on 50 A-MCMC runs. We partition the waveform (see Fig.~\ref{fig:totalwaveform}) into 50  sections, and for each section we calculate the periodogram and the energy spectral density of the SGWB in the context of \textit{LISA} noise and astrophysical background. We estimate the parameters of the model of Eq.~\ref{eq:modelPSDs} using Eq.~\ref{eq:DWDLF} for the amplitude of the DWD SGWB at low frequencies. 

With this method, we measure the spectral energy density's amplitude of the Galactic foreground at different periods of the year. 

The observed modulation is an effect of the \textit{LISA}  measurement with  the \textit{LISA} antenna pattern. For each section $i$, we assume that the \textit{LISA} pattern antenna is constant. The waveform of the section $i$ is:
\begin{equation}
    s_i(t) = h_{+,i}(t) F_{+,i} + h_{\times,i}(t) F_{\times,i}.
\end{equation}
The spectral energy density of the waveform $s_i$ becomes:
\begin{equation}
    \begin{split}
        S_{s_i}(f) &= \tilde{s}_i\tilde{s}^*_i(f) = \tilde{h}_{+,i}\tilde{h}^*_{+,i}(f) F^2_{+,i} +\tilde{h}_{\times,i}\tilde{h}^*_{\times,i}(f) F^2_{\times,i}\\
        & = S_h(f)\left(F^2_{+,i} + F^2_{\times,i} \right)\\
    \end{split}
\end{equation}
with $S_h(f) = 2 \tilde{h}_{+,i}\tilde{h}^*_{+,i}(f) = 2 \tilde{h}_{\times,i}\tilde{h}^*_{\times,i}(f)$. For each section we have the estimate of the modulated DWD amplitude at low frequencies; each section corresponds to a different part of the year. We can thus measure the orbital effect of the change over the year:

\begin{equation}\label{eq:fitMod}
        \Omega_{Mod,i} =\Omega_{DWD,LF}^u\left(F^2_{+,i} + F^2_{\times,i} \right)\\
\end{equation}
where $\Omega_{DWD,LF}^u$ is the unmodulated amplitude of the energy spectral density of the Galactic foreground at low frequencies. 
The error is given by the standard deviation estimated from the posterior distributions of the chains. 

\begin{figure*}
\centering
\includegraphics[height=22cm]{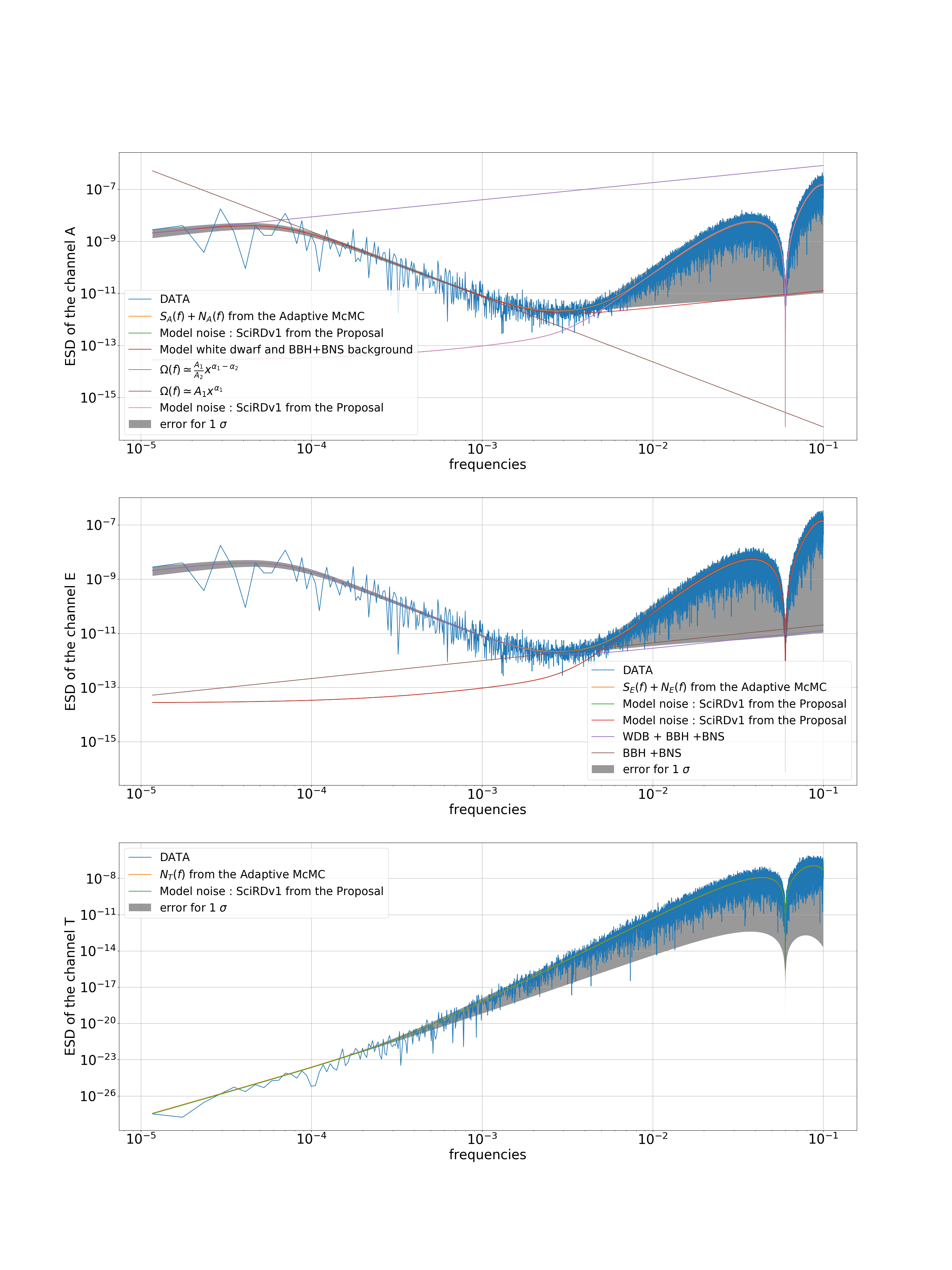}
\caption{Energy spectral density for the channels $A, E$, and $T$ from section 30 of the 50. Estimates are made for the two magnitudes of the \textit{LISA} noise model from the proposal~\citep{2019arXiv190706482B}, the modulation of the WD foreground and the astrophysical background, with an A-MCMC of 8 parameters (BBH + DWD + \textit{LISA} noise)}
\label{fig:Data+MCMC30}
\end{figure*}

\begin{figure*}
\centering
\includegraphics[height=17cm]{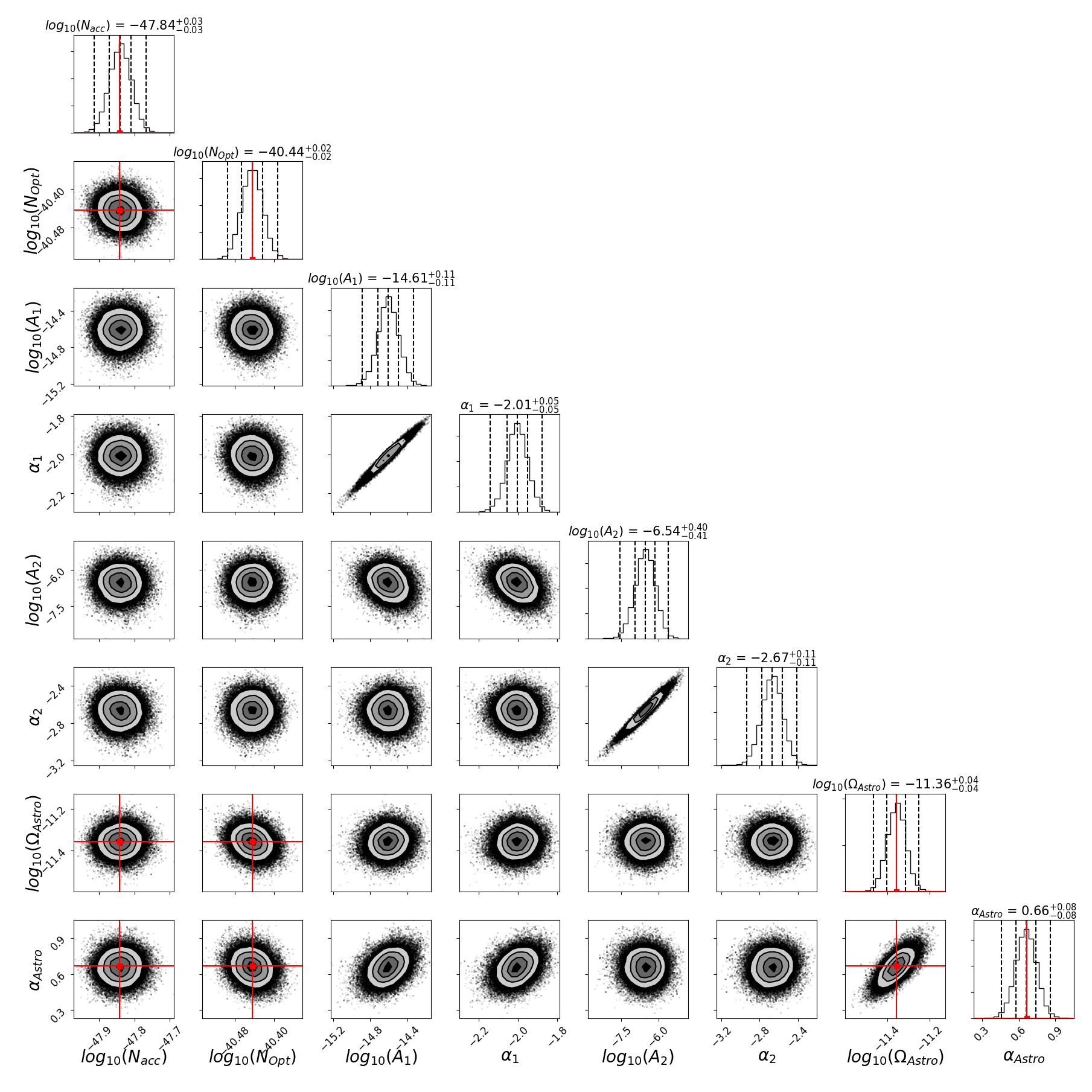}
\caption{Corner plot for the A-MCMC of 8 parameters (BBH/BNS + WD Section N$^{\circ}$30 + \textit{LISA} noise) using the channels $A$, $E$ and $T$. The results are for the two magnitudes for the \textit{LISA} noise model from the proposal \citep{2019arXiv190706482B}, the power law SGWB (amplitude and spectral slope), and the DWD foreground (two magnitudes and two slopes). The vertical dashed lines on the posterior distribution represent from left to right the quantiles $[16\%,\ 50\%,\ 84\% ]$. The red lines are the "true" parameter values.}
\label{fig:corner30}
\end{figure*}

As an example, in Fig.~\ref{fig:Data+MCMC30} we display the results from on the 50 A-MCMC runs, namely section 30 of the orbit. The blue lines are the three periodograms for the data channels $A, E,$ and $T$ with $N_{acc}, N _{Pos} = \bigg(1.44 \ 10^{-48} \ \text{s}^{-4}\text{Hz}^{-1}$, $3.6 \ 10^{-41} \ \text{Hz}^{-1}\bigg)$. We estimate the two \textit{LISA} noise magnitudes and the 6 stochastic background parameters $(A_1, \alpha_1, A_2, \alpha_2)$ for the DWD and $(\Omega_{GW,astro}, \alpha_{astro})$ for the astrophysical background. We use log uniform priors with five magnitudes for the five amplitude parameters $(N_{acc}, N_{Pos}, A_1,  A_2, \Omega_{GW,astro})$, and  uniform priors for the slopes.
The A-MCMC parameters are set to $\beta = 0.01$, $N =  200 \ 000$, and we use 100 samples to estimate the covariance matrix.
The orange lines represent the \textit{LISA} noise and the energy spectral density of the SGWBs; see Eq.~\ref{eq:modelPSDs}.   The green lines are the results of the A-MCMC, and in grey the 1 $\sigma$ errors. Fig.~\ref{fig:corner30} shows the corner plot of the A-MCMC orange line in the Fig~\ref{fig:Data+MCMC30}; the marginal posterior distributions are symmetric. 

It is also possible to have very efficient estimates of the different noise components thanks to the signal $T$ being nominally devoid of GW signals. We have verified that over the one year of data, and the 50 A-MCMC results, the parameters are constant, except for the parameter $A_1$, which varies due to the orbital modulation.

\begin{figure*}
\centering
\includegraphics[height=22cm]{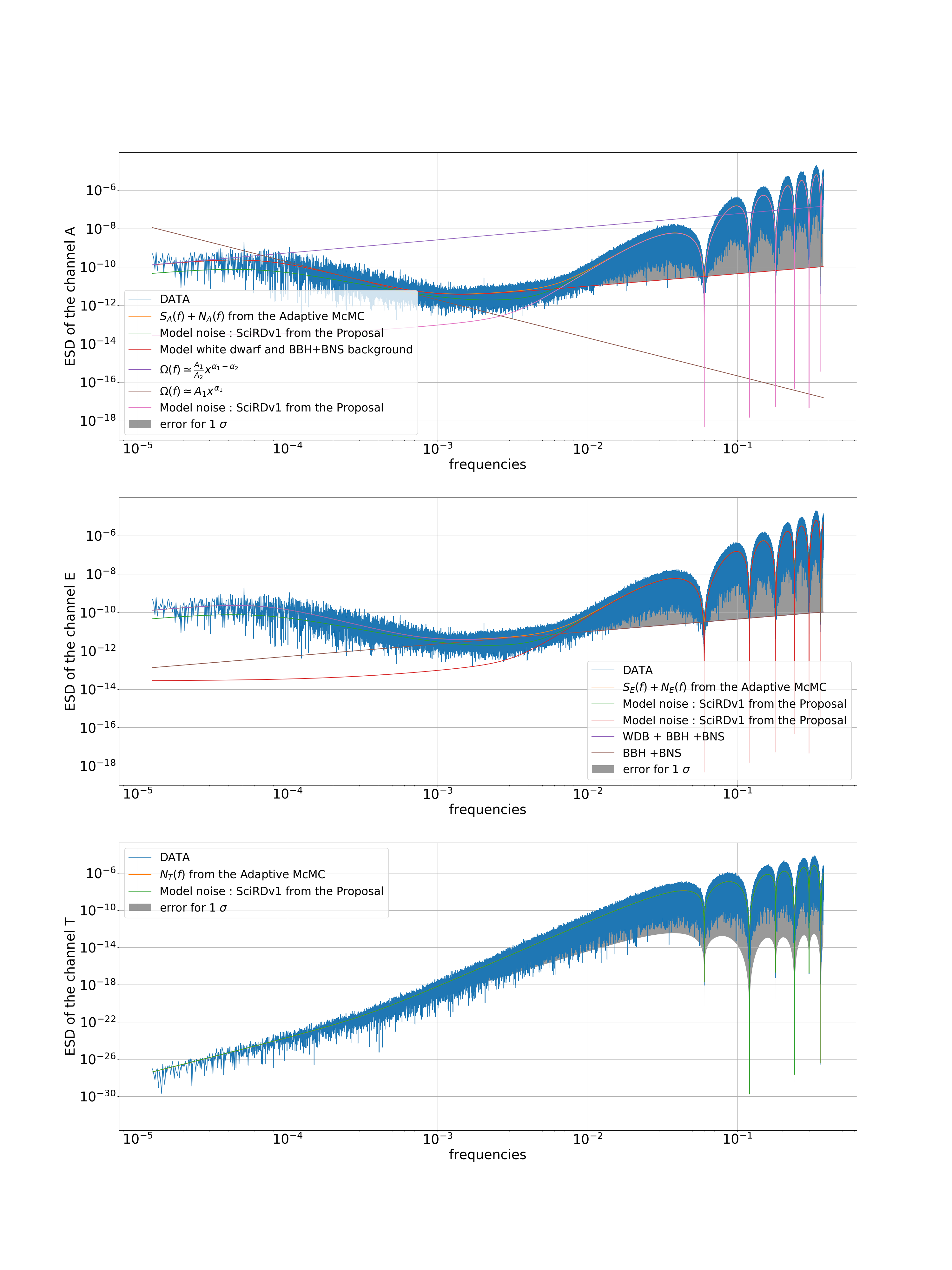}
\caption{Energy spectral density for the channels $A, E$, and $T$ from the total waveform. Estimates are made for the two magnitudes of the \textit{LISA} noise model from the proposal~\citep{2019arXiv190706482B}, the modulation of the WD foreground, and the astrophysical background with an A-MCMC of 8 parameters (BBH + WD + \textit{LISA} noise). One year of data was used.}
\label{fig_corncovmcmc}
\end{figure*}

\begin{table}
	\centering
	\caption{Run N$^{\circ}$30 of the 50 A-MCMC run parameter estimation results for the year of data with a GW background input of $\Omega_{GW,astro}= 4.4 \times 10^{-12}\left( \frac{f}{3 \times 10^{-3} \ \text{Hz}} \right)^{2/3}$, and the Galactic DWD binary foreground from \citet{10.1093/mnras/stz2834}. In Fig.~\ref{fig:corner30}, the red lines correspond to the true values.}
	\label{tab:mcmctable30}
	\begin{tabular}{lcccr} 
		\hline
		& $\Omega_{GW,astro}$ & $\alpha_{astro}$ & $\Omega_{DWD,LF}$  & $\alpha_{DWD,LF}$ \\
		\hline
		$\mu$    & $4.38 \ \times 10^{-12}$ & $0.66$ & $2.01 \ \times 10^{-9}$ & $0.66$  \\
		$\sigma$ & $2.35 \ \times 10^{-13}$  & $0.08$ & $1.34 \ \times 10^{-9}$ & $0.1$  \\
\hline
\end{tabular}
\label{tab:gal-astro30}
\end{table}

\begin{table}
	\centering
	\caption{50 A-MCMC run parameter estimation results for the year of data with a GW background input of $\Omega_{GW,astro}= 4.4 \times 10^{-12}\left( \frac{f}{3 \times 10^{-3} \ \text{Hz}} \right)^{2/3}$, and the Galactic DWD binary foreground from \citet{10.1093/mnras/stz2834}. In Fig.~\ref{fig_cornervmcmc}, the red lines correspond to the true values.}
	\label{tab:mcmctable}
	\begin{tabular}{lcccr} 
		\hline
		& $\Omega_{GW,astro}$ & $\alpha_{astro}$ & $\Omega_{DWD,LF}$  & $\alpha_{DWD,LF}$ \\
		\hline
		$\mu$    & $4.46 \ \times 10^{-12}$ & $0.65$ & $5.67 \ \times 10^{-9}$ & $0.68$  \\
		$\sigma$ & $1.2 \ \times 10^{-13}$  & $0.04$ & $3.89 \ \times 10^{-9}$ & $0.06$  \\
\hline
\end{tabular}
\label{tab:gal-astro}
\end{table}

At low frequencies, the model of the broken power law of DWD energy spectral density (see Eq~\ref{eq:newmodel}) can be approximated by a power law function, for a WD binary foreground the slope $\alpha = \alpha_1-\alpha_2 = \frac{2}{3}$.
For $1 \ll A_2 \left(\frac{f}{f_{ref}}\right)^{\alpha_2}$ (low frequency: LF):
\begin{equation} \label{eq:DWDLF}
     \Omega_{DWD,LF}(f) \approx \frac{A_1}{A_2} \left(\frac{f}{f_{ref}}\right)^{\alpha_1-\alpha_2} 
\end{equation}

Presented in Fig.~\ref{fig:std_waveform} with the dashed green line is an estimate of the modulated DWD amplitude of Eq.~\ref{eq:fitMod} from the 50 A-MCMC runs. We measure the  amplitude of the energy spectral density of the Galactic foreground at low frequency $\Omega_{DWD,LF}^u=(3.9\pm 1.14 )\times 10^{-9}$ for a reference frequency $3 \times 10^{-3} \ \text{Hz}$. We use the \texttt{scipy.optimize.curve\_fit} method of least squares to fit Eq.~\ref{eq:fitMod} to estimate the amplitude of the Galactic foreground at low frequencies \citep{2020SciPy-NMeth}. The input of the least squares approximation is the modulated DWD amplitude at low frequencies (see Eq.~\ref{eq:DWDLF}). We also use the estimated standard deviation from the 50 A-MCMC runs as an input to the least squares procedure, with the argument \texttt{sigma} set to the error from 50 A-MCMC. This result corresponds to the 'real' measurement of the Galactic foreground amplitude without the modulation. The brown line represents the spectral separability for the Bayesian study of the energy spectral density of the Galactic foreground with the low frequency limit for the total waveform length of one year. 

The mean value of the 1500 estimates of $\Omega_{Mod}$ can be seen in blue, which is the mean of the grey curve. For the Bayesian analysis, the brown cross is no better, as no conclusive information appears. As can be seen with the error-bar, one cannot properly estimate the Galactic foreground amplitude without accounting for the changing \textit{LISA} antenna response. 

We have a good understanding of the signal modulation from the resolved binaries \citep{2014PhRvD..89b2001A}, and from theory. By identifying the resolvable foreground, one can predict the level of the unresolvable background. We note that the method presented in \citet{2014PhRvD..89b2001A} is more powerful, but for our present study the method we use is sufficient as an input to study the limitation of the measurement of the cosmological SGWB.
We have consistent results with our algorithm. Indeed, we correctly estimate the limit at low frequencies, and moreover, the astrophysical background is also estimated accurately, with less than 3 \% error. We have more difficulty fitting the Galactic foreground (68 \% error); this is due to the low frequency adjustment. 
 We note too that in Fig.~\ref{fig:std_waveform} the second peak at 0.75 year is higher than the first at 0.25 year. This detail has been also noted by \citeauthor{PhysRevD.71.122003} and \citeauthor{2014PhRvD..89b2001A}.

\begin{figure*}
\centering
\includegraphics[height=17cm]{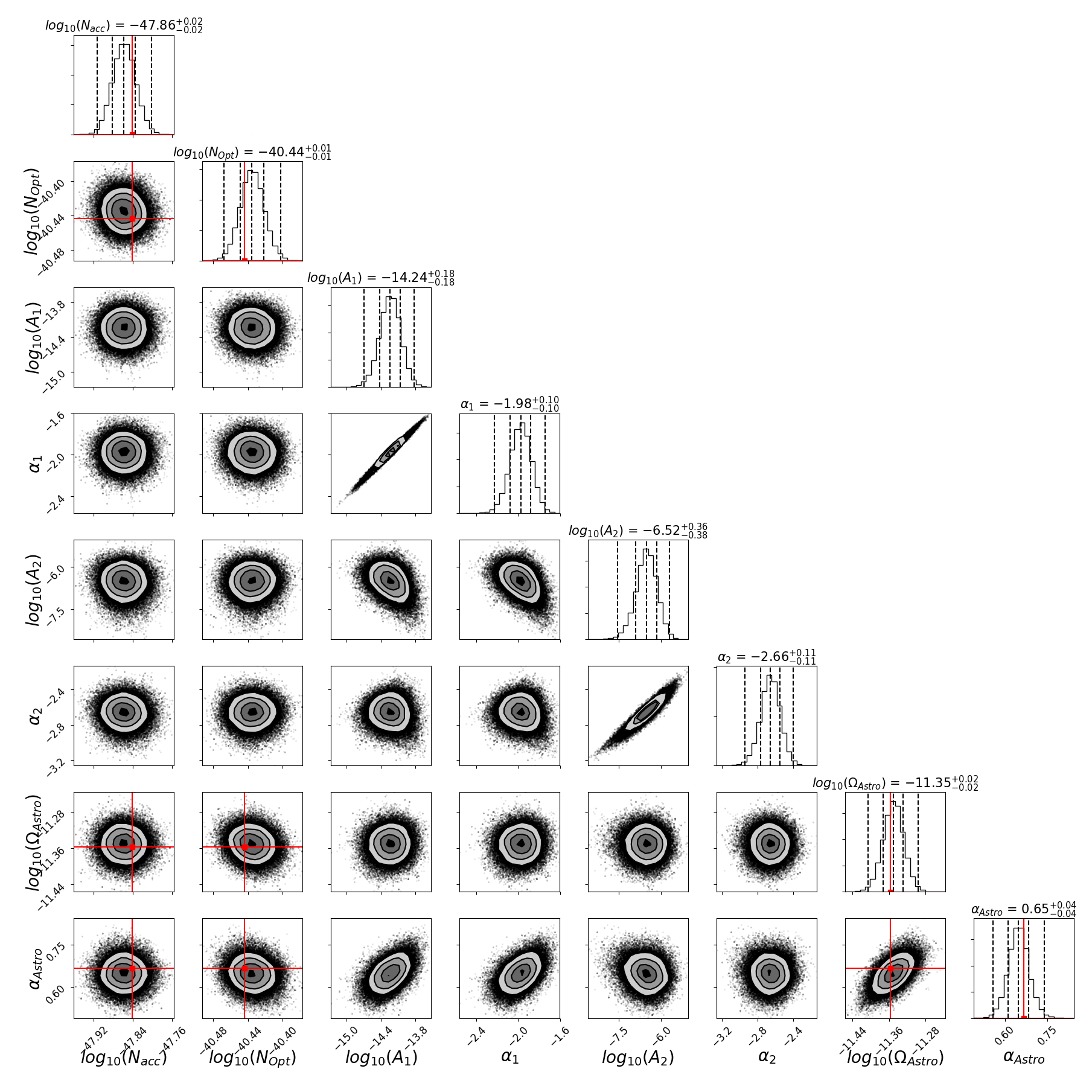}
\caption{Corner plot for the A-MCMC of 8 parameters (BBH/BNS + WD + \textit{LISA} noise) using the channels $A$, $E$ and $T$ with one year of data. The results are for the two magnitudes for the \textit{LISA} noise model from the proposal \citep{2019arXiv190706482B}, one single power law SGWB (amplitudes and spectral slopes) and the DWD (two magnitudes and two slopes). The vertical dashed lines on the posterior distribution represent from left to right the quantiles $[16\%,\ 50\%,\ 84\% ]$. The red lines are the "true" parameter values.}
\label{fig_cornervmcmc}
\end{figure*}

From the year of data, and the 50 A-MCMC results, the parameter estimates and errors for the Galactic foreground and astrophysical background are presented in Table~\ref{tab:gal-astro} and Table~\ref{tab:gal-astro30}. 
Fig~\ref{fig_corncovmcmc}
shows the energy spectral density estimates and Fig~\ref{fig_cornervmcmc} displays the corner plots  of all model parameters; these results were generated using a full year of data.
This demonstrates that \textit{LISA} can successfully observe and describe an astrophysically produced background from compact binaries, a Galactic DWD foreground, and \textit{LISA} detector noise and separate these  SGWB components.

\section{Measurement of the Cosmological SGWB}\label{sc:cosmolimitation}

In this section, we present the goal of our study, namely the ability for \textit{LISA} to measure a cosmological background in the presence of other stochastic signals and noise. 

\citet{2020arXiv201105055B} presented the evidence for the separability of the cosmological and astrophysical backgrounds with a precision around $\Omega_{GW, cosmo} \approx 1 \times 10^{-12} $ to $1 \times 10^{-13} $.

As indicated in Section~\ref{sc:cov}, it is possible to estimate the measurement error for each parameter using the Fisher matrix. Eq.~\ref{eq:fisher} gives the Fisher matrix, which depends on the parameters to be estimated, and also on the data collection time. Indeed, we assume that \textit{LISA} noise is a zero mean random noise, and that it is independent of the GW signal that we are trying to measure. 

The SGWB signal from year to year is essentially the same. 
By integrating the data over time one can reduce the influence of the \textit{LISA} noise on the SGWB search. We use the following for the magnitudes of the \textit{LISA} noise: acceleration noise of the test-mass  $N_{acc} = 1.44 \times 10^{-48} \ \text{s}^{-4} \text{Hz}^{-1}$; and optical metrology system noise $N_{Pos} = 3.6 \times 10^{-41} \ \text{Hz}^{-1}$). 
For the Fischer matrix study we consider observation times of  1, 4, 6, and 10 years. Thus, we will be able to see the effect of the integration of time in attempting to measure the cosmological background. We calculate the measurement uncertainty of the magnitude of the  cosmological background for several mission durations and for cosmological normalised energy densities between $1\times 10^{-14}$ and $1\times 10^{-8}$. We set a limit on the ability to detect a cosmological SGWB. We calculate the uncertainty of the measurement of the amplitude of the cosmological background. If this uncertainty is less than 50\%, we claim that the background is detectable and separable from the \textit{LISA} noise, the Galactic foreground and the astrophysical SGWB. Above this limit, it is impossible to conclude on the presence or not of a cosmological SGWB.

\begin{figure*}
\centering
\includegraphics[height=22cm]{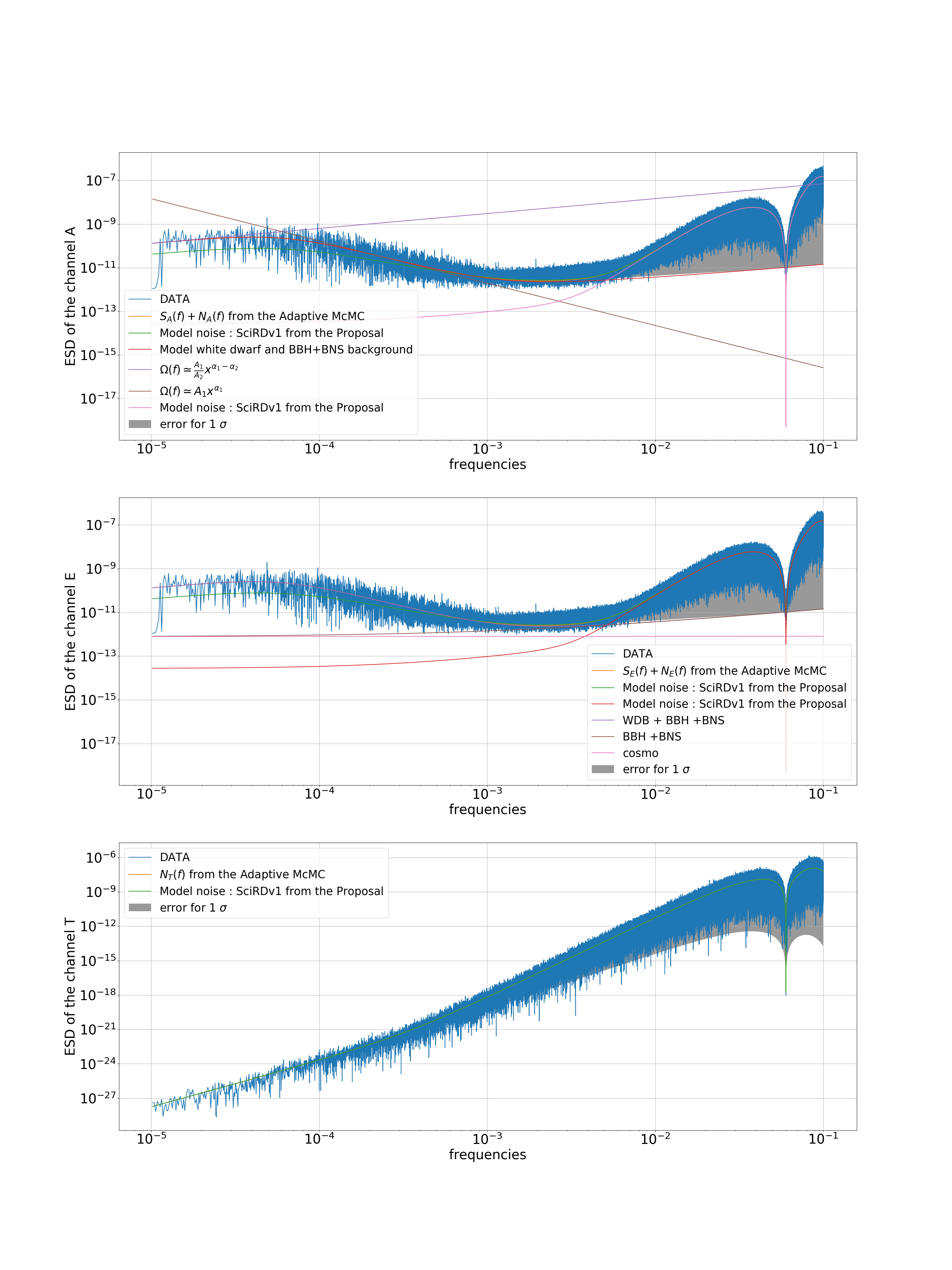}
\caption{Energy spectral density for the channels $A, E$, and $T$ from the total waveform of 1 year data seen by \textit{LISA}. Estimates are made for the two magnitudes of the \textit{LISA} noise model from the proposal~\citep{2019arXiv190706482B}, the modulation of the WD foreground, the astrophysical background and a cosmological background with an A-MCMC of 10 parameters (cosmological + BBH + DWD + \textit{LISA} noise)}
\label{fig:Data+MCMC}
\end{figure*}

\begin{figure*}
\centering
\includegraphics[height=17cm]{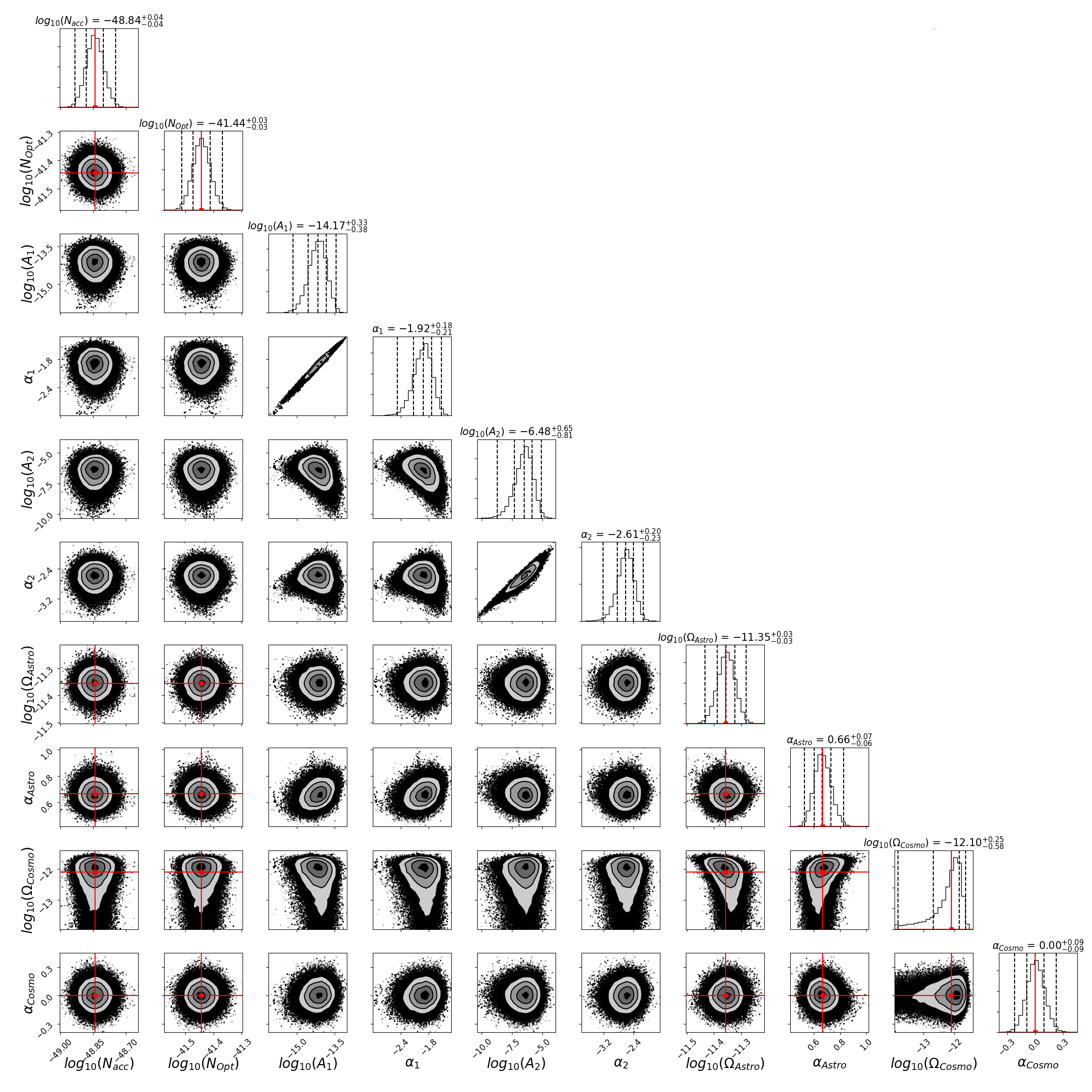}
\caption{Corner plot for the A-MCMC of 10 parameters (BBH/BNS + WD (total waveform of 1 year data seen by \textit{LISA}) + cosmo + \textit{LISA} noise) using the channels $A$, $E$ and $T$. The results are for the two magnitudes for the \textit{LISA} noise model from the proposal \citep{2019arXiv190706482B}, and two power law SGWB (amplitudes and spectral slopes) and the DWD (two magnitudes and two slopes). The vertical dashed lines on the posterior distribution represent from left to right the quantiles $[16\%,\ 50\%,\ 84\% ]$. The red lines are the "true" parameter values.}
\label{fig:corner}
\end{figure*}

We also conduct a Bayesian study using an A-MCMC algorithm to estimate the parameters of our model: two magnitudes for the \textit{LISA} noise; two parameters for the cosmological background (amplitude and slope); two parameters for the astrophysical background (amplitude and slope); and four parameters for the broken power law (two amplitudes and two slopes).
In all, we estimate 10 parameters based on the three periodograms from channels $A$, $E$, and $T$.  We use the astrophysical background from \citet{2019ApJ...871...97C}. We vary the amplitude of the cosmological background to determine the precision with which it can be detected. Thus, we can produce parametric estimates using the A-MCMC for cosmological normalised energy densities injected with levels between $ 1 \times 10^{-14} $ and $ 1 \times10^{-8}$, all with a slope of $\alpha_{cosmo} = 0$.
In  Fig.~\ref{fig:Data+MCMC},  the blue lines are the three periodograms for the data channels $A, E,$ and $T$ for one year of data simulated  with $N_{acc},N_{Pos} = \bigg(1.44 \ 10^{-48} \ \text{s}^{-4}\text{Hz}^{-1}$, $3.6 \ 10^{-41} \ \text{Hz}^{-1}\bigg)$. We estimate the two \textit{LISA} noise parameters and the 8 GW parameters $(A_1, \alpha_1, A_2, \alpha_2)$ for the DWD, $(\Omega_{astro}, \alpha_{astro})$ for the astrophysical background and $(\Omega_{cosmo}, \alpha_{cosmo})$ for the cosmological background. This is an example of the separability with input comprising the Galactic binaries from \citeauthor{10.1093/mnras/stz2834} and the astrophysical SGWB $\Omega_{astro}=4.4 \times 10^{-12}$ at $3 \times 10^{-3} \ \text{Hz}$ with a slope of $2/3$ and a flat ($\alpha_{cosmo} = 0$) cosmological SGWB $\Omega_{cosmo}=8 \times 10^{-13}$. The orange lines represent the model used, see Eq.~\ref{eq:modelPSDs}. The A-MCMC is characterized by $\beta = 0.01$, $N =  200 \ 000$, and we use 100 MCMC samples to estimate the co-variance matrix. We use log uniform priors with six magnitudes for two LISA noise magnitudes and the four SGWB amplitude parameters $(N_{acc},N_{Pos},A_1, A_2,\Omega_{astro}, \Omega_{cosmo})$ and a uniform prior for the slopes $(\alpha_1,\alpha_2, \alpha_{astro}, \alpha_{cosmo})$ (2 degrees of freedom). The green lines are the results of the A-MCMC, and in grey the errors for 1 $\sigma$. Fig.~\ref{fig:corner} displays the corner plot for all parameters based on one year of data; the posterior distributions are symmetric. We have the evidence for a good fit for the astrophysical background and the cosmological background.

Table~\ref{tab:mcmctable10} is the summary of the results with the cosmological input $\Omega_{cosmo}= 8 \times 10^{-13}$. This is just at the level of detectability for the cosmological background.  A year of data was used, and the results come from the 50 A-MCMC results. This shows that it will be possible for \textit{LISA} to distinguish the cosmological background at this level from the astrophysical background, the Galactic DWD foreground, and \textit{LISA} detector noise.

\begin{table*}
	\centering
	\caption{Results of the year long 50 A-MCMC runs  with an input SGWB of $\Omega_{astro}= 4.4 \times 10^{-12}\left( \frac{f}{3 \times 10^{-3} \ \text{Hz}} \right)^{2/3}$, and a cosmological input $\Omega_{cosmo}= 8 \times 10^{-13}$ with a slope $\alpha_{cosmo}= 0$. We use as reference frequency $3 \times 10^{-3} \ \text{Hz}$. Also presented are the low frequency (LF) results for the Galactic DWD. In Fig.~\ref{fig:corner} the red lines correspond the true values.}
	\label{tab:mcmctable10}
	\begin{tabular}{lcccccr} 
		\hline
		& $\Omega_{astro}$ & $\alpha_{astro}$ & $\Omega_{cosmo}$ & $\alpha_{cosmo}$ & $\Omega_{DWD,LF}$  & $\alpha_{DWD,LF}$ \\
		\hline
		$\mu$    & $4.41 \ \times 10^{-12}$ & $0.67$  & $8.01 \ \times 10^{-13}$ & $0.04$ & $5.88 \ \times 10^{-9}$ & $0.68$  \\
		$\sigma$ & $1.74 \ \times 10^{-13}$  & $0.06$ & $3.97 \ \times 10^{-13}$  & $0.09$ & $6.5 \ \times 10^{-9}$ & $0.13$  \\
\hline
\end{tabular}
\end{table*}

\begin{figure*}
\centering
\includegraphics[height=9cm]{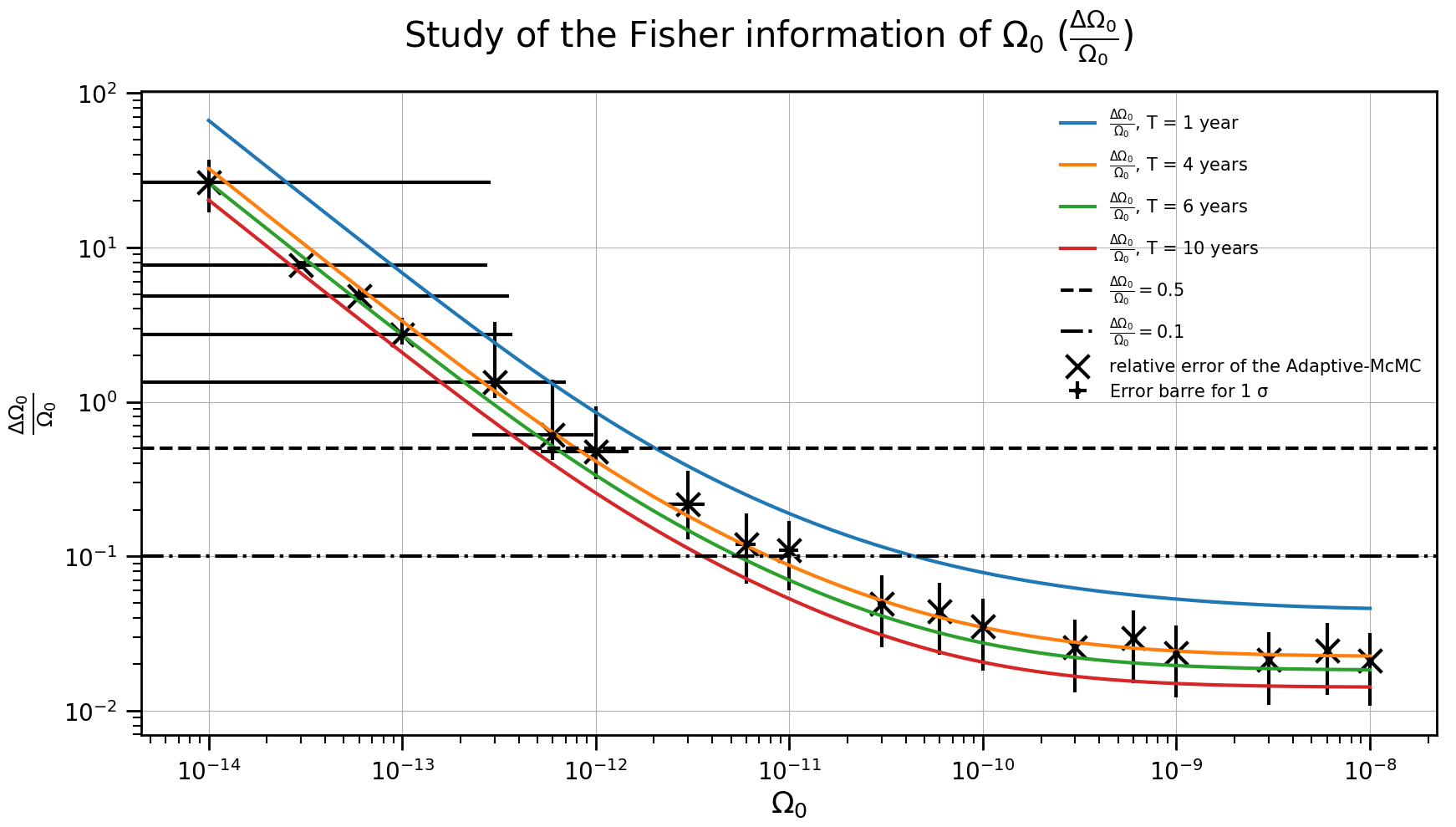}
\caption{Uncertainty estimation of the cosmologically produced SGWB from the Fisher information study (represented as the colored lines for 4 different observational time durations) and the parameter estimation from the A-MCMC (in scatters) for the channels $A$ and $E$ with the noise channel $T$. The upper horizontal dashed line represents the error level $50\%$. Above the line the error is greater than $50\%$.}
\label{fig:Uncertainty}
\end{figure*}

Fig.~\ref{fig:Uncertainty} displays the uncertainties for the measurement of the cosmological background as we vary its amplitude. 
We assume a flat background, with
$\alpha_{cosmo} = 0$ and $\Omega_{cosmo} = \Omega_{0}$.
The results from two studies are presented.  The first is the Fisher matrix study, presented as lines of  blue, orange, green and red, corresponding to \textit{LISA} observation durations of 1, 4, 6 and 10 years. We see that the effect of the duration does not have a large influence. Indeed, we explain this by noting the frequency dependence of the noise in the periodogram, which is predominantly at high frequencies, but we measure the GW backgrounds essentially at low frequencies. Despite this, we can see that the temporal dependence is not zero. A longer integration time allows a better fit. Our Bayesian study (see the black scatter), of which we present the results from A-MCMC runs for one year of data.  Each point 
has an error bar obtained by estimating the standard deviation of the posterior distribution. 
Clearly, the measurement uncertainty is greater for low amplitudes of the cosmological SGWB.
There is a very good agreement between the A-MCMC results and the Fisher matrix analysis.
With our detection criterion, $ \frac{\Delta \Omega_{0}}{\Omega_{0}} <0.5$, 
we can predict that with our method is it is possible to fit an SGWB of cosmological origin of $\Omega_{0,lim} = 8 \ \times 10^{-13}$, given the values we have used for the \textit{LISA} noise, the galactic foreground and the astrophysical background.

\subsection{Fisher Information Studies for modified Galactic Foreground Models}

The galactic foreground as well as the astrophysical background are very uncertain. In our first study we considered different levels for the compact binary produced astrophysical background ($\alpha = 2/3$), in the range of $\Omega_{\textrm{GW,astro}}(25 ~ \textrm{Hz}) = 0.355 \rightarrow 35.5 \times 10^{-9}$; with this we showed that it would be possible with LISA to measure a cosmologically produced SGWB ($\alpha = 0$) in the range of $\Omega_{\textrm{GW,cosmo}} \approx 1 \times 10^{-12}$ to $1 \times 10^{-13}$ with 4 years of observation~\citep{2020arXiv201105055B}. 

\begin{figure*}
\centering
\includegraphics[height=9cm]{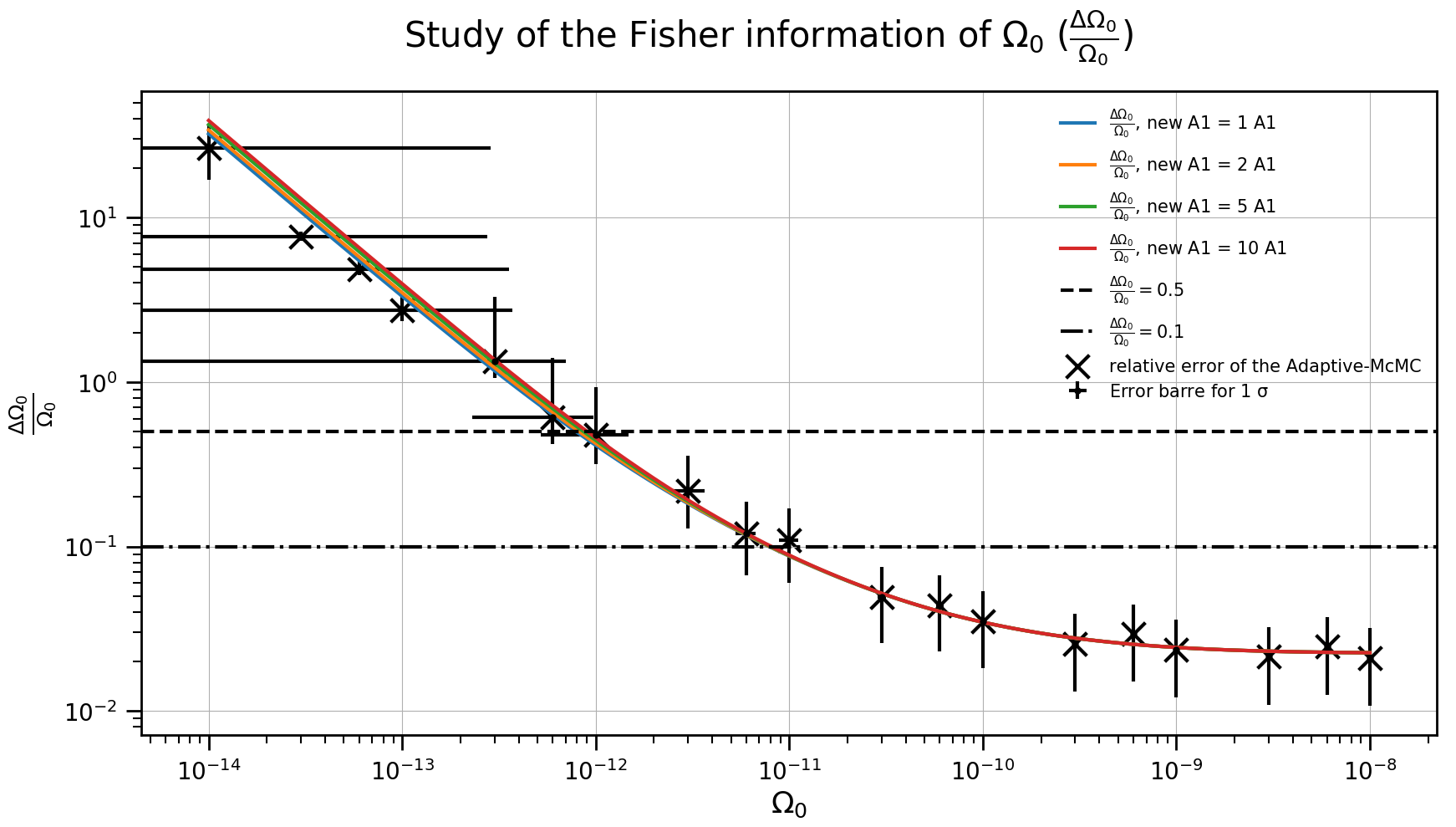}
\caption{Uncertainty estimation of the cosmologically produced SGWB from the Fisher information study (represented as the colored lines for 4 different amplitudes $A_1$) and the parameter estimation from the A-MCMC (in scatters) for the channels $A$ and $E$ with the noise channel $T$. The upper horizontal dashed line represents the error level $50\%$. Above the line the error is greater than $50\%$. The parameter representing the magnitude of the galactic foreground has been modified, namely $A_{1,new} = d\times A_1 $ for $d=1,2,5,10$; see Eq.~\ref{eq:newmodel}}
\label{fig:Uncertainty2}
\end{figure*}

Now we address the uncertainty in the DWD galactic model.
In this subsection, we investigate the effect of modifying the density model for the galactic foreground. We test the influence of modifications with a Fisher information study. Indeed, in this paper, we have shown that the two studies (Fisher and A-MCMC) give very similar results. First we introduce a modification of the amplitude of the galactic foreground by testing the separability for new forced values of the parameter $A_1$, such as $A_{1,new} = d\times A_1 $ for $d=1,2,5,10$; see Eq.~\ref{eq:newmodel}. Fig.~\ref{fig:Uncertainty2} presents the uncertainty for the cosmologically produced SGWB normalized energy density with the variation of the amplitude $A_1$. Increasing the parameter $A_1$ only slightly decreases the possibility of measuring a SGWB of cosmological origin. This is a modification at very low frequencies $f_ {GW} < 7.2 \times 10^{-5}$, and does not significantly influence the estimation of the cosmological background by LISA.  

\begin{figure*}
\centering
\includegraphics[height=9cm]{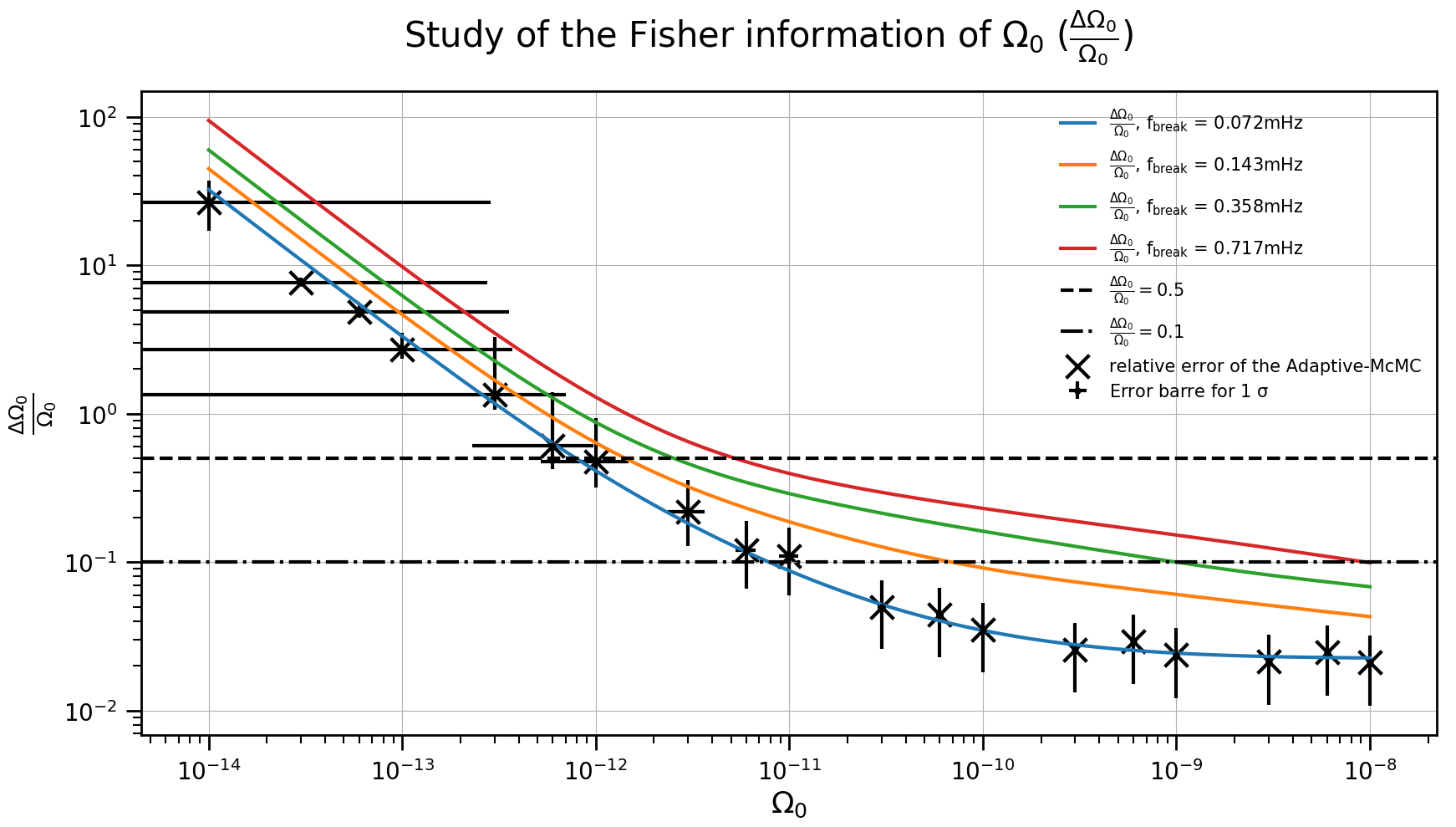}
\caption{Uncertainty estimation of the cosmologically produced SGWB from the Fisher information study (represented as the colored lines for 4 different peak frequencies in changing the spectral dependence $f_{break}$) and the parameter estimation from the A-MCMC (in scatters) for the channels $A$ and $E$ with the noise channel $T$. A factor $d$ defines a new frequency of separation of the two spectral dependencies of the galactic foreground, $f_{new,break} = d \times f_{break}$. We test $d = 1,2,5,10$.
The upper horizontal dashed line represents the error level $50\%$. Above the line the error is greater than $50\%$.}
\label{fig:Uncertainty1}
\end{figure*}

We also introduce a modification of the frequency position of the zone of influence of the two spectral dependencies for the two slopes of our broken power law. It is possible to show that the cutoff frequency $f_{break}$ is given by $f_{break} = f_{ref}e^{\frac{-\ln(A_2)}{\alpha_2}} $. The change in frequency is given by a modification of the amplitude $A_2$ (again, see Eq.~\ref{eq:newmodel}), such that the new amplitude is given by $A_{2,new} = A_2 d^{-\alpha_2}$, where $d$ is the multiplying coefficient giving the new frequency of separation of the two spectral dependencies of the galactic foreground ($f_{new,break} = d \times f_{break}$). We conduct the Fisher information study for $d = 1,2,5,10$. In Fig.~\ref{fig:Uncertainty1} we show the uncertainty of the cosmological SGWB estimation for different $f_{break}$ values.  We note that a spectral shift towards the higher frequencies, decreases our possibility of measuring the cosmological background. For a value of $d = 10$ the limit of detection is increased to $\Omega_{0} \approx 6 \times 10^{-12}$. This is logical because the shift of the galactic foreground to higher frequencies would more strongly affect the measurement of the cosmological background. 

It is important to note that the galactic foreground will not be just DWDs, it may also contain WD-M-dwarf (White dwarf + M dwarf binaries; a M dwarf can also be called a red dwarf), stripped stars or CVs. WD-M-dwarfs are few in comparison to DWDs, furthermore, we estimate that they are very low frequency objects so if we consider them we would expect a very slight increase in the $A_1$ parameter, which does not change our results \citep{Skinner_2017}. From Fig. 2 of \citet{G_tberg_2020}, we see that binary stripped stars are less numerous than DWDs, they are also present at very low frequency. Moreover from Fig. 3 of \citet{G_tberg_2020} the chirp mass is most important. So, adding this population also modifies the $A_1$ parameter, which should not change our conclusion. There are likely too few CVs to generate a significant high frequency foreground~\citep{Meliani:2000gz,Marsh_2011,vanderSluys:2011yc,Pala2020_CVGaia} most of the high amplitude and high frequency CV sources will produce resolved events. 
Another important LISA signal source are AM Canum Venaticorum (AM CVn) stars. AM CVn binaries have been observed with periods between 5 and 65 minutes, hence gravitational-wave sources from $5 \times 10^{-4}$ to $7 \times 10^{-3}$ Hz~\citep{vanderSluys:2011yc}. Much is still unknown about their space density from observations~\citep{Carter_2012,Carter_2013} and theoretical studies~\citep{Nelemans:2004pp,Kremer:2017xrg,Breivik:2017jip}, and estimates of the space density can vary by several orders of magnitude.
Clearly the importance of understanding the binaries in the Milky Way will be meaningful for LISA searches, including the SGWB.

\section{Conclusions}
\label{sec:conclusion}
This study has displayed what may be possible for \textit{LISA} in its ability to observe a cosmologically produced SGWB in the presence of an astrophysical BBH produced background, a Galactic DWD foreground, and inherent \textit{LISA} detector noise. This paper also presents a comparison between two DWD catalogs~\citep{10.1093/mnras/stz2834,Nelemans_2001}. 

We find that the positional distribution of DWD does change the shape of the energy spectrum of the Galactic foreground. Our study can be easily applied to other catalogs of DWDs. For preparations of the \textit{LISA} SGWB observations, it will be important to consider models that are as close as possible to the real Galactic distribution. 
\textit{LISA} will have the ability to observe Galactic DWDs, both with resolvable binaries and the stochastic foreground, and make important statements about the distribution in the Galaxy. In this present study
we do not observe significant changes to the GW power spectrum between resolved and unresolved DWDs. This is also the case for the different compositions of the cores of the WDs. We have studied the distribution of resolved and unresolved binaries according to their core compositions. It does not seem possible to us to extrapolate the chemical composition of all the binaries with the resolved binaries.

Our analysis considered the distribution of DWD produced GW signals in the Galaxy, and the detection response by \textit{LISA} as it orbits the sun, rotates its configuration, and changes it orientation with respect to the Galaxy. A modulation of the observed Galactic DWD foreground appears. Accurate parameter estimation for the different SGWB backgrounds (astrophysical, cosmological) must accurately estimate the signal modulation and amplitude from the Galactic foreground. Building on previous analyses addressing the modulated signal from the Galaxy~\citep{PhysRevD.71.122003,2014PhRvD..89b2001A}, we have presented a strategy to demodulate and measure the spectral energy density of the Galactic foreground at low frequencies. The orbital modulation of the Galactic foreground aids in the parameter estimation for the isotropic (and hence unmodulated) astrophysical and cosmological SGWBs.

We show that it will be possible to measure the SGWB amplitude of cosmological origin $\Omega_{GW,cosmo} \approx 8 \ \times 10^{-13}$  with an error of less than 50\%. In our study, we consider this SGWB to have flat spectral energy density $\propto f^0$ \citet{Cornish_2001}; we note that this is an approximation for more complex cosmological backgrounds.  Phase transition in the early universe can produce two-part power laws, with a traction between the rising and falling power law components at some peak frequency; a more complex version of our algorithm should be able to perform parameter estimation for these types of backgrounds as well. In our present study, the cosmological background prediction is obtained with an astrophysical background estimated to be at a level consistent with the observations made by Advanced LIGO and Advanced Virgo \citep{2019ApJ...871...97C}. It is important to note that this astrophysically produced SGWB is the main source of limitation for \textit{LISA} in its effort to observe a cosmologically produced SGWB. An extragalactic background from DWDs could increase complexity as well.

Future third-generation projects, the Einstein telescope~\citep{Punturo_2010} or Cosmic Explorer~\citep{Reitze:2019iox}, will also be trying to observe a cosmologically produced SGWB in the presence of an astrophysically produced background. However, these third generation detectors operating at higher frequencies, above 5 Hz,
could have such detection sensitivity that almost all binary black hole mergers in the observable universe would be directly observable~\citep{Regimbau:2016ike}, and then could be removed from the SGWB search. The first consequence is therefore the disappearance of the astrophysical SGWB from the study of separability. So according to \citet{PhysRevD.102.024051} the ability to detect the cosmological background will be further improved.

\section*{Acknowledgements}
\label{sec:awk}
GB, AL, NC and NJC  thank the Centre national d'études spatiales (CNES) for support for this research. 
NJC appreciates the support of the NASA \textit{LISA} Preparatory Science grant 80NSSC19K0320. 
RM acknowledges support by the James Cook Fellowship from Government funding, administered by the Royal Society Te Ap\={a}rangi and
DFG Grant KI 1443/3-2.


\section*{Data Availability}

The data generated for our study presented in this article will be shared upon reasonable request to the corresponding author.



\bibliographystyle{mnras}
\bibliography{example} 

\bsp	
\label{lastpage}
\end{document}